\documentclass[graphicx]{aa}
\usepackage{xcolor}
\usepackage{ulem}
\newcommand{\varv}{v}
\newcommand{\beq}{\begin{equation}}
\newcommand{\beqa}{\begin{eqnarray}}
\newcommand{\eeq}{\end{equation}}
\newcommand{\eeqa}{\end{eqnarray}}
\def\vec#1{\ensuremath{\mathchoice{\mbox{\boldmath$\displaystyle#1$}}
{\mbox{\boldmath$\textstyle#1$}}
{\mbox{\boldmath$\scriptstyle#1$}}
{\mbox{\boldmath$\scriptscriptstyle#1$}}
}}
\begin{document}
\title{Magnetically threaded accretion disks in resistive magnetohydrodynamic simulations and asymptotic expansion}
%   \subtitle{}
\titlerunning{Magnetically threaded accretion disks}
\authorrunning{M. \v{C}emelji\'{c} et al.}
\author{M. \v{C}emelji\'{c},
\inst{1,2,3,*}
W. Klu\'{z}niak
\inst{2,1}
\and
V. Parthasarathy
\inst{4}
 }
%\offprints{M. \v{C}emelji\'{c}}
\institute{Research Centre for Computational Physics and Data Processing, Institute of Physics, Silesian University in Opava, Bezru\v{c}ovo n\'am.~13, CZ-746\,01 Opava, Czech Republic
\and
Nicolaus Copernicus Astronomical Center, Polish Academy of Sciences,
Bartycka 18, 00-716, Warsaw, Poland
\and
Academia Sinica, Institute of Astronomy and Astrophysics, P.O. Box 23-141,
Taipei 106, Taiwan
\and
H\"ochstleistungsrechenzentrum Stuttgart, Nobelstra\ss e 19, 70569
Stuttgart, Germany\\
\\
$^*$\email{miki@camk.edu.pl}}
\date{Received ??; accepted ??}
\abstract
{}
{A realistic model of magnetic linkage between a central object and its
accretion disk is a prerequisite for understanding the spin history of stars and stellar remnants. To this end, we aim to provide an analytic
model in agreement with magnetohydrodynamic (MHD) simulations.
}
{For the first time, we wrote a full set of stationary asymptotic expansion equations of a
thin magnetic accretion disk, including the induction
and energy equations. We also performed a resistive MHD
simulation of an accretion disk around a star endowed with a magnetic
dipole, using the publicly available code PLUTO. We compared the analytical results with the numerical solutions, and
discussed the results in the context of previous solutions of the induction
equation describing the star-disk magnetospheric interaction.}
{We found that the magnetic field threading the disk is suppressed by orders of magnitude inside thin disks, so the presence of the stellar magnetic field does not strongly affect the velocity field, nor the density profile inside the disk. Density and velocity fields found in the MHD simulations match the radial and vertical profiles of the analytic solution.
Qualitatively, the MHD simulations result in an internal magnetic field
similar to the solutions previously obtained by solving the induction
equation in the disk alone. However, the magnetic field configuration
is quantitatively affected by magnetic field inflation outside the disk; this is reflected in the net torque. The torque on the star is an order of magnitude larger in the magnetic than in the non-magnetic case. Spin-up of the star occurs on a timescale comparable to the accretion timescale in the MHD case, and is an order of magnitude slower in the absence of a stellar magnetic field.
}
   {}


\maketitle

\section{Introduction}
\label{intro}
While the problem is more generally applicable to accreting stellar
systems, the question of the interaction between the stellar
magnetosphere and the accretion disk has gained particular prominence
in discussions of spin-up and spin-down torques of X-ray pulsars. It
is not controversial that the accretion disk is truncated at an inner
radius where the influence of the stellar magnetosphere balances the
stresses in the disk. The exact location of this inner radius,
$r_\mathrm{in}$, depends on the model, but most expressions differ by
no more than a factor of two.\footnote{For a review refer to the Appendix
in \cite{KRapp2007}.}

There is less agreement on the detailed disk-magnetosphere interaction
and the resulting torques on the star. One school of thought assumes
the disk to be perfectly conducting. \cite{Scharlemann1978} and
\cite{Aly1980} considered the interaction of the accretion disk and
the stellar magnetic field to be concentrated only in a very narrow
ring near the inner edge of the disk. Simulations of the magnetosphere
in the presence of a perfectly conducting disk seem to be in agreement
with those ideas \citep{Romanova2012,parf1,parf2,parf3}.

In another paradigm, the idealization of a perfectly conducting plasma
is given up, and the stellar magnetic field is allowed to penetrate a
broader region of the inner accretion disk through turbulent diffusion
and reconnection \citep{Vasyliunas1975, Ghosh1977,Ghosh1979}, leading
to the classic picture of magnetic torques contributing to spin-up in
the region $r_\mathrm{in}<r<r_{\rm cor}$ and to spin-down in the region
$r>r_{\rm cor}$, where $r_{\rm cor}=(GM_\star/\Omega_\star^2)^{1/3}$ is
the co-rotation radius, $M_\star$ is the mass of the star, and
$\Omega_\star$ its rotation rate. The resulting torques in a modified
model within this paradigm were computed by \cite{Wang1996}, and the
associated changes in the structure of the accretion disk by
\cite{KRapp2007}; c.f. \cite{Agapitou,Alpar}. A recent paper
discusses the problem in the context of ultraluminous X-ray pulsars
for a modified disk solution, with the inclusion of advection
\citep{Chashkina2019}. All the aforementioned analytic work assumed a
thin height-integrated disk, i.e., followed the \citet{ss73} scheme.

A more fundamental approach within this main-stream paradigm has been
taken by \cite{naso1,naso2}, who actually solved the induction equation
inside the accretion disk, on a fixed background of plasma variables
corresponding to the \citet{ss73} $\alpha$-disk model. \cite{naso3}
have performed the same task, but on the background of the \citet{KK00} model,
computing the torque contribution from the interior of the disk (and
not the disk surface alone, as was done by previous authors) and
finding that different vertical zones of the disk may contribute to
spin-up or spin-down regardless of whether they are inside or outside
the co-rotation radius.

We would like to take the next step and to solve simultaneously and 
self-consistently for both the magnetic field and the fluid variables,
such as the velocity field, in a magnetically threaded disk. To this
end, we have developed a very stable numerical solution of an accretion
disk, and have performed resistive MHD simulations of the accretion
disk interacting with the stellar magnetosphere. It is numerically
expensive to perform such simulations when the stellar magnetic
dipole becomes too strong. For this, and other reasons it would be
desirable to have arguments allowing extension of the results obtained
to the more strongly magnetized regime. An analytic solution would
serve this purpose admirably, but is rather hard to obtain. As a
first step we will be satisfied with finding in an analytic
calculation certain restrictions on the solution, as well as scaling
laws which validate the MHD numerical solution, thus allowing an
extension of the latter to stronger magnetic fields.

In this paper we present both our numerical resistive MHD simulation
solutions to the disk-magnetosphere problem, as well as an asymptotic
expansion of the governing equations which yield constraints on the
magnetic field in leading orders of the dimensionless disk thickness,
and allow a partial analytic solution of the problem. Our numerical simulations include the innermost part of the accretion disk and accretion column quite close to the stellar surface. While we find the formal Alfv\'{e}n radius (defined as the radius at which the Alfv\'{e}n and radial flow velocities become equal) to be typically about 4 stellar radii in the disk, the flow starts to lift from the disk, following the magnetic field lines to the pole, only at less than 3 stellar radii. This truncation radius of the disk also happens to be close to the co-rotation radius for our simulations. 
We compare our analytical solutions with the numerical ones at 6 and 15 stellar radii in the disk.

In the following, after an introduction to thin disks (\S~\ref{thin}),
in \S~\ref{asympt} we present the equations that are solved in the work,
and outline the results of the method of asymptotic expansion. We then
present the quasi-stationary results in our numerical simulations
(\S~\ref{numcomp}) and find the expressions for the best matches to
the numerical solutions. In \S~\ref{torq} we discuss the torques on the star,
and summarize the results in \S~\ref{concl}. Appendix A contains a
detailed analysis of the equations by the method of asymptotic expansion, and Appendix B contains details of the numerical setup.

%%%%%%%%%%%%%%%%%%%%%%%%%%%%%%%%%%%%%%%%%%%%%%%%%%%%%%%%%%%%%%%%%%%%%%%%%%
\section{Thin disks}
%%%%%%%%%%%%%%%%%%%%%%%%%%%%%%%%%%%%%%%%%%%%%%%%%%%%%%%%%%%%%%%%%%%%%%%%%%
\label{thin}
Stationary disks are in vertical hydrostatic equilibrium. We restrict
ourselves to geometrically thin disks with the height scale $H$ much smaller than the radial distance $r$, $H<<r$, whence the radial
gravitational attraction of the central body is balanced by the
(differential) rotation of the disk. For such disks this rotation very
closely follows the Keplerian law. It is not clear whether thin disks can
extend to the direct vicinity of a (weakly magnetized) neutron star at
near-Eddington accretion rates, as the inner disk would then be
radiation-pressure dominated and subject to thermal
instability\footnote{While stable solutions have recently been
found in this regime, they are not geometrically thin
\citep{Lancova2019}.} on the Keplerian timescale
\citep{ss76,Mishra1,Sadowski16,Mishra2}. However, at lower accretion
rates, or further away from the star (as in the strongly magnetized
classic X-ray pulsars), the disk is gas-pressure dominated and
stable. Our simulations and analytic results correspond to the
gas-pressure dominated regime.

The first analytical solution for the accretion disk has been given
in \citet{ss73} in the $\alpha$ approximation, in which the stress
tensor is proportional to the pressure. In that and many following
works the radial solution was obtained as an average over the disk
thickness, with equations in the vertical direction separately solved
to obtain a hydrostatic balance. However, \citet{urp84} showed that
height-averaged values of radial velocity are not representative of
the disk midplane, where in fact the flow direction is directed
radially outwards. 

A three-dimensional, analytic solution for the velocity field
and density of a steady axisymmetric, polytropic, hydrodynamical (HD)
$\alpha$-disk has been given in \citet{Kita95} and \citet[]{KK00},
hereafter KK00, and confirmed in \citet{Reg02} with the use of the
ideal equation of state. Equatorial backflow is present for all
realistic values of the viscosity parameter ($\alpha<0.68$). These
results were upheld by the results of several numerical simulations
\citep[e.g.,][] 
{Igumen,William,MishraR20a,MishraR22}; and also \cite{phiraf17}.

As remarked in the Introduction, most of the work on the
disk-magnetosphere problem was performed for \cite{ss73} axially
symmetric $\alpha$-disks, which by their one-dimensional nature do
not allow backflows. To avoid the backflow, in part for greater ease
of comparison with previous work, and in part because we are
considering stationary MHD solutions and we are uncertain that
such solutions actually exist\footnote{After the submission of this
work, \cite{MishraR22} have shown that for Prandtl numbers below a
certain critical value backflow exists also in $\alpha$-disks
simulated with resistive MHD.} for $\alpha$-disks with backflows,
in this paper we assume the value $\alpha=1$ for the viscosity
parameter.

We extend the KK00 solution to the case of a magnetic disk. Since the
solution inside the disk depends on details of the star-disk
magnetospheric interaction, we cannot solve for the magnetic field in
the disk without knowing the global solution. For this reason, we
perform an MHD solution of the disk initially penetrated by the dipole
magnetic field of the star, and find the resulting velocity field, as
well as the magnetic field configuration, which exhibits the well-known
effect of field inflation. However, we will show analytically that the
vertical density profile of the magnetized thin accretion disk is the
same as in a hydrodynamic (HD) disk, and in the steady regime its
velocity field is not affected in the
leading order by the presence of the magnetic field. This is borne out 
by our simulations. Thus, for $\alpha=1$ the configuration of the field
internal to the disk is qualitatively similar to that found by
\cite{naso2}, who solved the induction equation on a background flow
solution corresponding to the \cite{ss73} solution.

%++++++++++++++++++++++++++++++++++++++++++++++++++++++++++++++++++++++
\begin{figure*}
\includegraphics[width=\columnwidth]{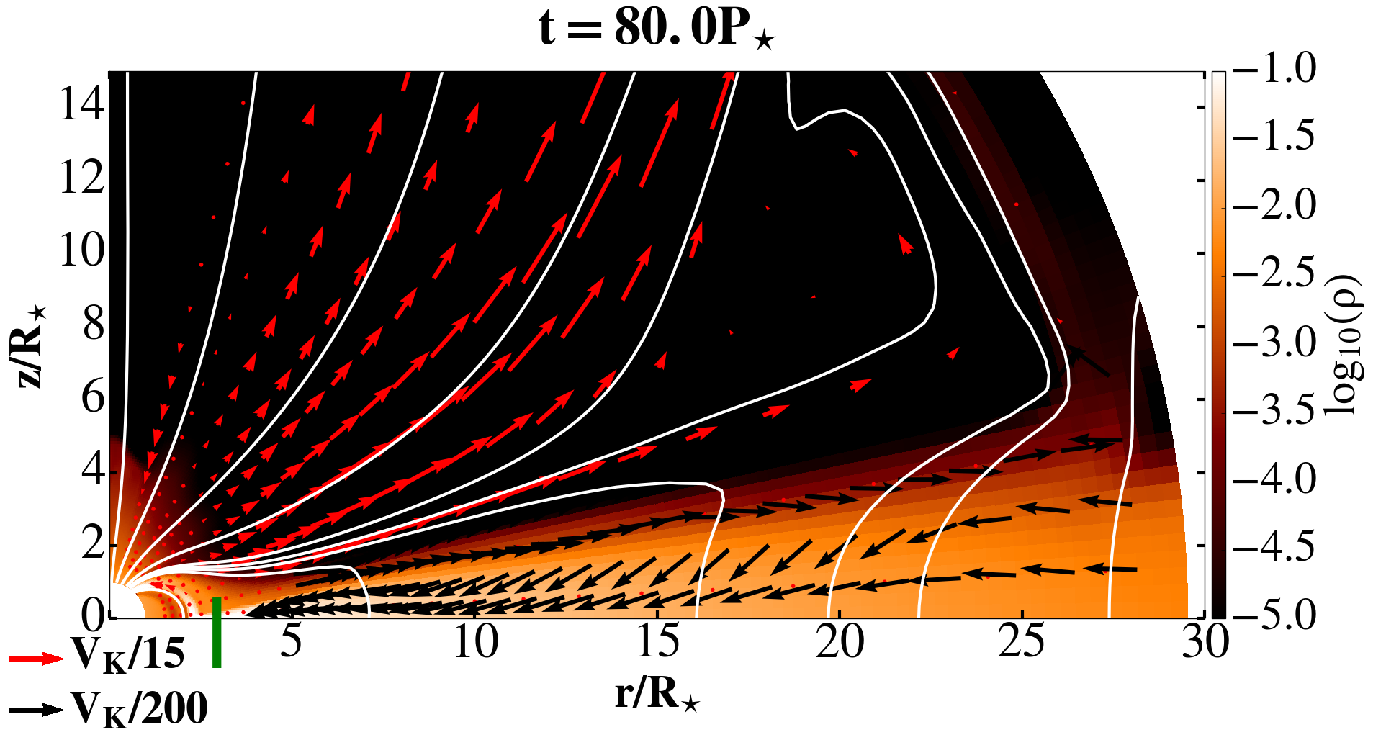}
\includegraphics[width=\columnwidth]{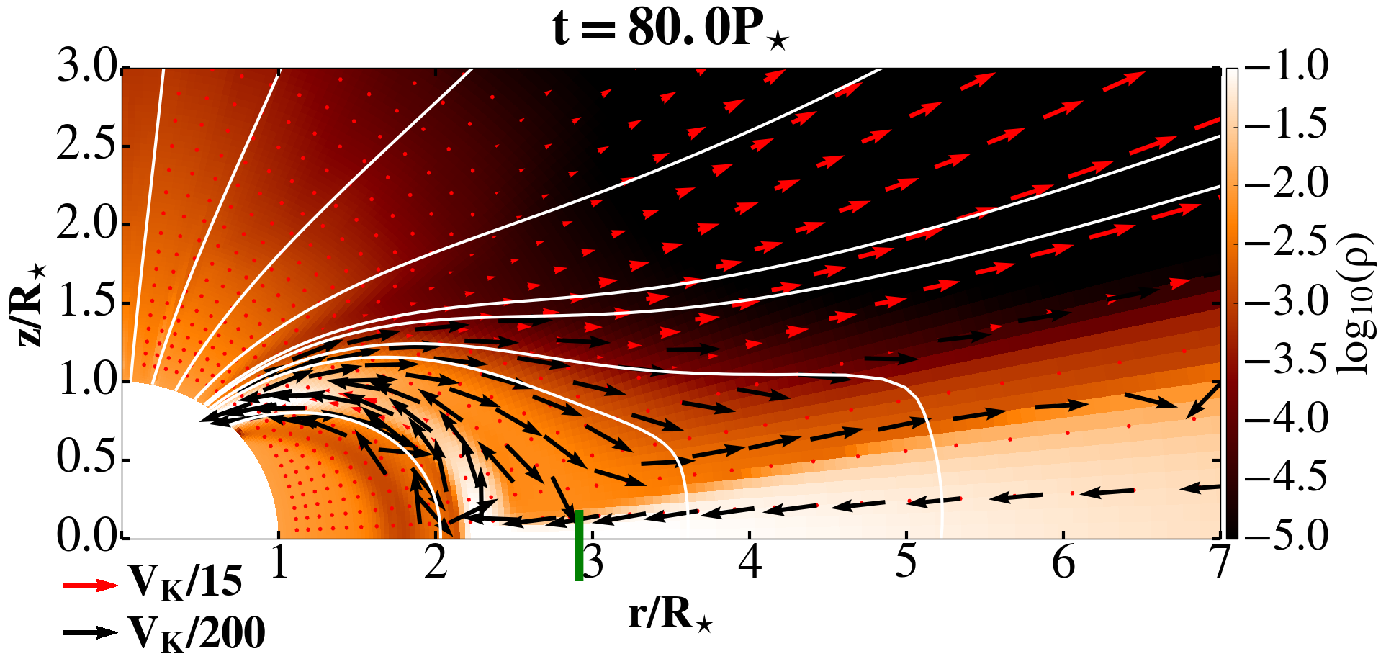}
\caption{Appearance of the inner disk in a zoomed snapshot in our
magnetic simulation after T=80 stellar rotations. The matter, is lifted
from the disk into the accretion column, feeding it onto the stellar
surface. The matter density is shown in logarithmic color grading in code
units, with a sample of velocity vectors shown in two normalizations: since
the velocity in the corona is much larger than in the disk, velocity vectors
are shown with a different scaling in the disk and the corona, the units of the respective velocity vectors being shown in the legend at the bottom left of the figures. With the
white solid lines is shown a sample of the poloidal magnetic field lines.
The green solid line marks the position of the corotation radius.}
\label{fig:diskreg}
\end{figure*}
%++++++++++++++++++++++++++++++++++++++++++++++++++++++++++++++++++++++
An extensive set of our simulations was shown in \cite{cem19}, where
quasi-stationary solutions were obtained with various stellar rotation
rates, magnetic field strengths, and magnetic Prandtl numbers.
Here we focus on the case of millisecond accreting pulsars, where the
disk extends very close to the stellar surface, and it is only at less
than three stellar radii ($r<3R_\star$) that the flow begins to follow
the magnetic field lines onto the pole. For neutron stars the solution
would correspond to a magnetic moment of $\sim10^{26}$G cm$^3$
%$\sim10^{30}$G cm$^3$
(surface
field of $\sim10^8\,$G). In our simulations the stellar rotation rate is
at 20\% of the equatorial mass-shedding limit, $\Omega_\star=0.20~\Omega_\mathrm{ms}$.

We will confront the disk solutions from such simulations with the
requirements obtained from the analytical equations. To do this,
we match the numerical solutions in the disk with a set of functional
expressions which best describe our simulations.
%%%%%%%%%%%%%%%%%%%%%%%%%%%%%%%%%%%%%%%%%%%%%%%%%%%%%%%%%%%%%%%%%%%%%%%%%%
\section{Thin magnetic accretion disk in asymptotic approximation}
\label{asympt}
%%%%%%%%%%%%%%%%%%%%%%%%%%%%%%%%%%%%%%%%%%%%%%%%%%%%%%%%%%%%%%%%%%%%%%%%%%

We are solving the viscous and resistive equations of magnetohydrodynamics
which are, in the cgs system of units:
\beqa
\nabla\cdot\vec{\mathrm B}=0,\label{eq1}\\
\frac{\partial\rho}{\partial t}+\nabla\cdot(\rho\vec{\mathrm v}) =0,
\label{eq2}\\ 
\frac{\partial\vec{\mathrm B}}{\partial t}+\nabla\times(\vec{\mathrm B}
\times\vec{\mathrm v}+\eta_{\mathrm m}\vec{J}/c)=0, 
\\
\frac{\partial\rho\vec{\mathrm v}}{\partial t}+\nabla\cdot\left[\rho\vec{\mathrm v}
\vec{\mathrm v}+\left(P+\frac{\vec{\mathrm B}\cdot\vec{\mathrm B}}{8\pi}\right)\vec{\mathrm I}-
\frac{\vec{\mathrm B}\vec{\mathrm B}}{4\pi}-\vec{\tau}\right]=\rho\vec{g},\\
\frac{\partial E}{\partial t}+\nabla\cdot\left[\left(E+P+\frac{\vec{\mathrm B}
\cdot\vec{\mathrm B}}{8\pi}\right)\vec{\mathrm v}-\frac{(\vec{\mathrm v}
    \cdot\vec{\mathrm B})\vec{\mathrm B}}{4\pi}\right]=\rho\vec{g}\cdot\vec{\mathrm v},\label{eq5}
\eeqa
where $\rho$, $P$, $\vec{\rm v}$ and $\vec{B}$ are the density, pressure,
velocity, and magnetic field, respectively. The symbols $\eta_{\rm m}$ and
$\vec{\tau}$ represent the Ohmic resistivity and the viscous stress
tensor, respectively, with $\vec{\tau}=\eta\vec{T}$, where $\eta$ is
the dynamic viscosity and $\vec{T}$ is the strain tensor.

The  acceleration of gravity is $\vec{g}=-\nabla\Phi_{\mathrm g}$, with
the gravitational potential of the star with mass $M_\star$ equal to
$\Phi_{\mathrm g}=-GM_\star/R$. The total energy density
$E=P/(\gamma-1)+\rho(\vec{\rm v}\cdot\vec{\rm v})/2$, and the electric
current is given by Ampere's law
$\vec{J}/c=\nabla\times\vec{B}/(4\pi)$. We assume the ideal gas with an
adiabatic index $\gamma=5/3$, and polytropic index $n=3/2$ (as $\gamma\equiv 1+1/n$).

We search for quasi-stationary state solutions, assuming that all
the heat is radiated away locally by the disk, at the rate at which it
is being generated, as is appropriate in the thin-disk approximation. This justifies neglecting the dissipative viscous
and resistive heating terms in the energy equation, and obviates the
need to introduce a cooling term. We still solve the equations in the
non-ideal MHD regime with viscous dissipation, because the viscous term is present in the momentum equation, and the Ohmic resistive term in the
induction equation.

The equations are solved numerically with the publicly available code PLUTO \citep{m07}, as described in \S~\ref{numcomp}, and Appendix B.
In Fig.~\ref{fig:diskreg} we show one example of our numerical result,
with a close up view
in the close vicinity of the central object to show the accretion column.
From the footpoint of the accretion column in the disk to the stellar
surface, the flow is in the ideal MHD regime, with a ``frozen in''
magnetic field inside the star-disk magnetosphere.

Both the equations and the resulting flow pattern seem too complicated to attempt an analytic solution. However, we would like to recover analytically the main features of the flow inside the disk, neglecting the stellar magnetosphere, except as an outer boundary condition on the disk solution, and ignoring the accretion column and the associated winds.
Following the approach of KK00, where the solution for the viscous,
hydrodynamical case of the thin accretion disk was derived, we
analyze the equations for the magnetized, resistive accretion disk in
the asymptotic expansion. For the first time, we consider also the
energy equation in this context. This will help us to evaluate
if omitting the heating terms in the energy equation is consistent with
other equations. For comparison of the terms in the equations, the
equations have to be written in normalized units.

In the asymptotic approximation, pioneered in the context of accretion
disk by \cite{Reg83}, all the variables are written in the Taylor
expansion in the small coefficient $\epsilon$, representing the
dimensionless height of the disk $\epsilon=\tilde{H}/\tilde{R}<<1$,
with the tildes representing typical values, in this equation of
the disk height H at radius R \citep[as in the general discussion of asymptotic approximation in KK00 and also][]{um06}. We can expand each
variable X as $X=X_0+\epsilon X_1+\epsilon^2 X_2+\epsilon^3 X_3+\dots $
and then, for each equation separately, consider all terms of the
same order in $\epsilon$.

We note that Eq.~(1) involves the magnetic field components alone, while
Eq.~(2) only the fluid variables. All the other equations couple the
fluid variables with the magnetic field. Even in the the purely HD
case ($\vec{\mathrm B}=0$) the equations constitute a system of
coupled partial differential equations, for four variables (density and the three components of velocity) if one assumes a polytropic equation of state.  It is remarkable that it is possible to find an analytic solution to these equations in the HD case, and to derive clear-cut conclusions in the MHD case ($\vec{\mathrm B}\neq 0$), as
shown below.

In the case of a viscous HD disk, Eqs.~(2) and (4) have been
solved inside the disk (KK00). When a stellar magnetic field is
present, the solution in the disk cannot be separated from the star-disk
magnetosphere, because of the connection of the magnetic field in the
corona with the field in the disk. In addition, a solution in the
magnetosphere can itself be complicated by the reconnection events and
outflows, and a back-reaction from the disk. Nonetheless, the asymptotic expansion method allows a derivation of constraints on the magnetic field inside the disk which are valid regardless of the field configuration outside it.

For the reader's convenience, in Appendix A we give a step-by-step
example of the asymptotic expansion in the equation of continuity,
Eq.~(2), together with the dimensionless thin-disk form of the remaining
four equations of resistive MHD that we are solving,
%\S~\ref{asympt}, 
with a derivation of the conditions (obtained
mostly from the zeroth and first order in $\epsilon$) which the magnetic
field should satisfy.

For instance, in the MHD case, just as in the HD KK00 results, for the
lowest order terms in velocity we obtain Keplerian motion
\beq
v_{r0}=v_{z\,0}=0,\  \Omega_0 \equiv v_{\phi 0}/r =r^{-3/2},
\eeq
with the first order correction terms to the fluid variables vanishing
(with the exception of radial velocity)
\beq
c_{\rm s1}=\ v_{z1}=v_{\varphi 1}=\rho_1=0,\\
v_{r1}\neq0.
\eeq

In the HD case, the non-Keplerian velocity components ($v_{r1}, v_{z2},
v_{\varphi2}$) can be obtained from the second order equations of momentum
conservation, yielding the radial and azimuthal velocities, $v_{r}$,
$r\Omega$, and from the equation of continuity yielding the vertical
component of velocity, $v_{z}$ (KK00). Now, in the MHD case, the first two
of these equations are coupled to the magnetic field components and their
derivatives.

\subsubsection*{Order $\epsilon^{2}$:}
\begin{equation}
\begin{aligned}
2r\rho_0\Omega_0\Omega_2=\frac{3\rho_0}{2}\frac{z^2}{r^4}+n\rho_0\frac{\partial c_{s0}^2}{\partial r}-\frac{\partial}{\partial z}\left(\eta_0\frac{\partial v_{r1}}{\partial z}\right)\\
-\frac{2}{\gamma\tilde{\beta}}\left( B_{r0}\frac{\partial B_{r0}}{\partial r}+B_{z0}\frac{\partial B_{r1}}{\partial z}-\frac{B_{\varphi 0}^2}{r}\right)+\frac{1}{\gamma\tilde{\beta}}\frac{\partial B_0^2}{\partial r},
\label{radmom2a}
\end{aligned}
\end{equation}
  
\begin{equation}
\begin{aligned}
\frac{\rho_{0}\varv_{r1}}{r}\frac{\partial}{\partial r}\left(r^{2}\Omega_0\right)= \frac{1}{r^2}\frac{\partial}{\partial r}
\left(r^3\eta_0\frac{\partial \Omega_0}{\partial r}\right)
+\frac{\partial}{\partial z}\left(\eta_0\frac{\partial\Omega_2}{\partial z}\right) \\
+\frac{2}{\gamma \tilde{\beta}}\left(B_{r0}\frac{\partial B_{\varphi 0}}
{\partial r}+B_{z0}\frac{\partial B_{\varphi 1}}{\partial z}+\frac{B_{r0}B_{\varphi
0}}{r}\right).
\label{Azmom2}
\end{aligned}
\end{equation}
We find (Appendix A) that the zeroth order components of the magnetic
field all vanish, $B_{i0}=0$, $i=r,z,\varphi$, with
$B_0^2\equiv B_{r0}^2 + B_{z0}^2 + B_{\varphi0}^2$ also zero.
On the assumption---well borne out by our simulations---that magnetic pressure is not dominant, $\mathcal{O}(\bar\beta)\ge1$, the equations for the fluid variables through second order are
exactly the same as for the HD case:
\beq
2r\rho_0\Omega_0\Omega_2=\frac{3\rho_0}{2}\frac{z^2}{r^4}+n\rho_0\frac{\partial c_{s0}^2}{\partial r}-\frac{\partial}{\partial z}\left(\eta_0\frac{\partial v_{r1}}{\partial z}\right),
\eeq
\beq
\frac{\rho_{0}\varv_{r1}}{r}\frac{\partial}{\partial r}\left(r^{2}\Omega_0\right)= \frac{1}{r^2}\frac{\partial}{\partial r}
\left(r^3\eta_0\frac{\partial \Omega_0}{\partial r}\right)
+\frac{\partial}{\partial z}\left(\eta_0\frac{\partial\Omega_2}{\partial z}\right),
\eeq
\begin{equation}
\frac{1}{r}\frac{\partial}{\partial r}\left(r\rho_0 v_{r1}\right)+
\frac{\partial}{\partial z}\left(\rho_0 v_{z2}\right) = 0.
\end{equation}
Clearly, with the equations being the same, any difference between the
HD and magnetic solution will depend on the boundary conditions. In
the magnetic case, it is the magnetic field at the disk surface and at its inner edge that is producing the difference in numerical solutions (between the
HD and MHD disks). We measure the overall effect of this difference by comparing
the initial HD setup with the quasi-stationary magnetic solution from our simulations.

Indeed, the vertical disk profile in \citet{H77} and KK00 has been
obtained by assuming that the disk density vanishes at the
surface, $\rho_0\rightarrow0$ as $z \rightarrow h$. If, instead, we supply at the disk
surface a value at the boundary with the coronal density $\rho_{\rm cd}$,
we obtain:
\beq
\rho_0(r,z)=\left[\rho_{\rm cd}^{2/3}(r)+\frac{h^2(r)-z^2}{5r^3}\right]^{3/2},
\label{rhozero}
\eeq 
where $h\equiv z_\mathrm{d}$ is the disk semi-thickness. The pressure and sound speed
now become:
\beq
P_0(r,z)=\left[\rho_{\rm cd}^{2/3}(r)+\frac{h^2(r)-z^2}{5r^3}\right]^{5/2},
\eeq
\beq
c_{\rm s{0}}(r,z)=
\sqrt{\frac{5}{3}\left[\rho_{\rm cd}^{2/3}(r)+\frac{h^2(r)-z^2}{5r^3}\right]}.
\label{pressound}
\eeq
Depending on the density contrast in the particular case under consideration,
one can estimate the value of $\rho_{\rm cd}$ and obtain the relevant solution.
The \citet{H77} solution is recovered by setting $\rho_{\rm cd}=0$.

In our case, since $h\propto r$, we can write, with the proportionality
constant $h'$,  $h=h'r$. Assuming the corona at the surface of the disk
to be in hydrostatic equilibrium, with
$\rho_{\rm cd}=\rho_{\mathrm c0}/r^{3/2}$ we can write:
\begin{equation}
\begin{aligned}
c_{\mathrm s0}^2(r,z)=\frac{5}{3}
\left(\rho_{\rm cd}^{2/3}(r)+\frac{h'^2r^2-z^2}{5r^3}\right)\\
=\frac{5}{3}
\left(\frac{\rho_{\mathrm c0}^{2/3}(r)}{r}+\frac{h'^2}{5r}-\frac{z^2}{5r^3}\right)\\
=\frac{\zeta^{-2}(r)}{3r}\left[1-\left(\zeta(r)\frac{z}{r}\right)^2\right]
=\frac{1}{3r^3}\left[\bar h^2(r)-{z}^2\right],
\end{aligned}
\label{cskvadr}
\end{equation}
and similarly for the pressure and density,
with 

\beq\zeta^{-2}=5\rho_{\mathrm c0}^{2/3}+h'^2.
\label{zeta}
\eeq
Thus,  the functional form of the vertical dependence of the density,
pressure, and speed of sound is the same as in the case of 
$\rho_{\rm cd}=0$, with
\beq
\bar h^2=h^2 + 5\rho_{\mathrm c0}^{2/3}r^2
\label{effh}
\eeq
playing the role of the effective height squared. 
%+++++++++++++++++++++++++++++++++++++++++++++++++++++++++++++++++++
\begin{figure}
\includegraphics[width=\columnwidth]{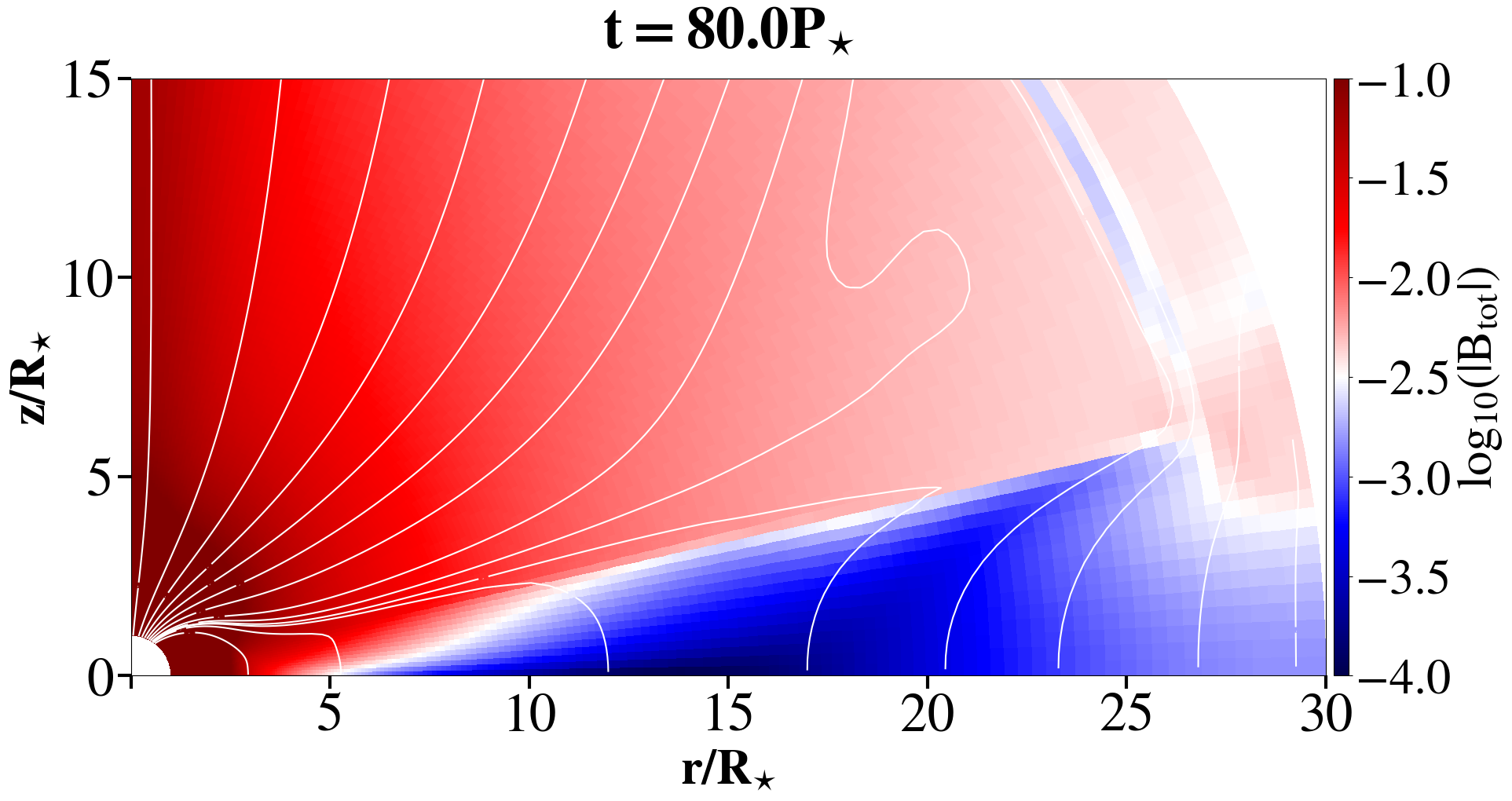}
\includegraphics[width=\columnwidth]{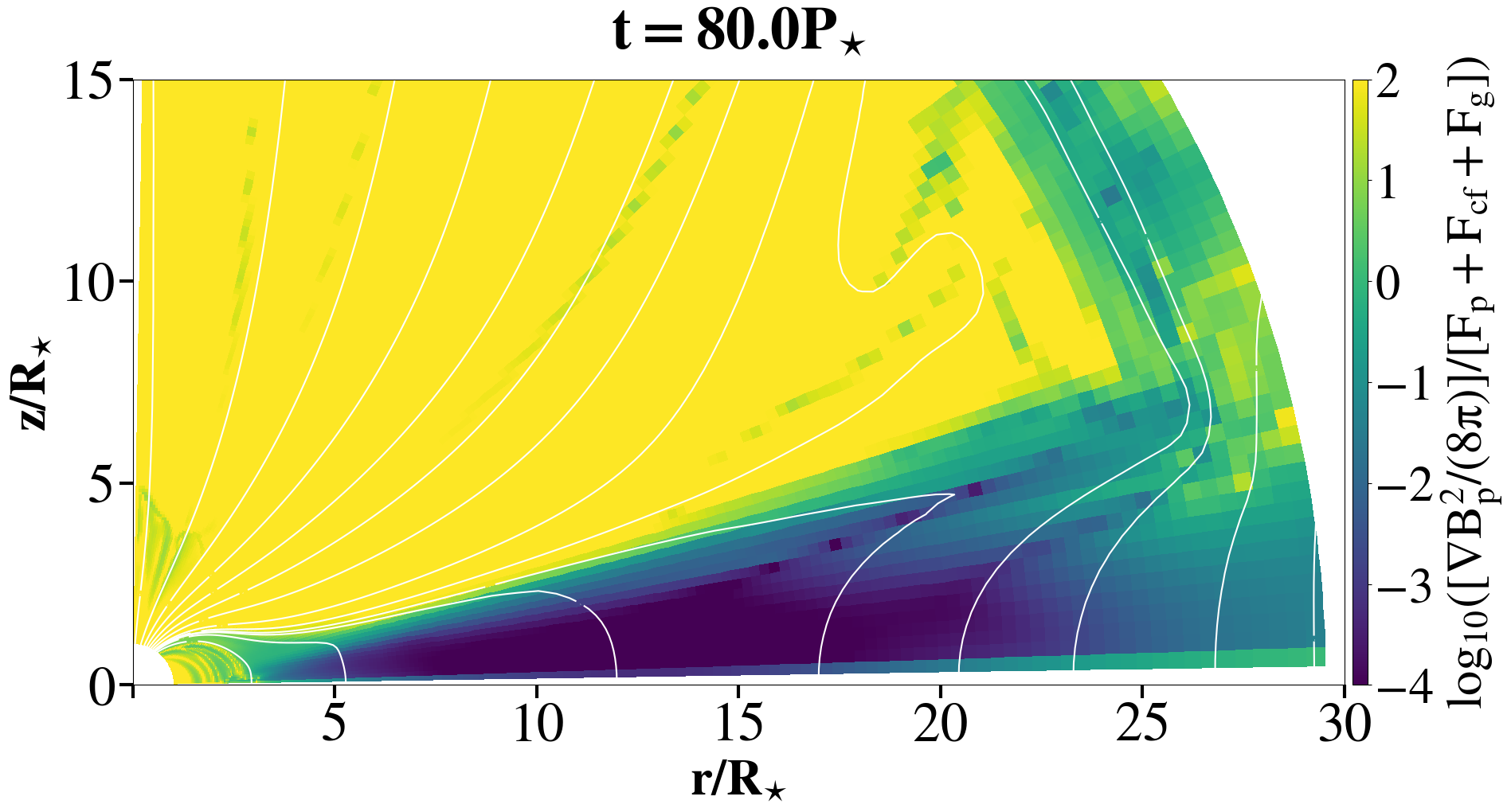}
\caption{{\it Top panel:} magnitude of the magnetic field in our simulation from Fig.~\ref{fig:diskreg}, with logarithmic
color grading in code units, and poloidal magnetic field lines shown with white solid lines. In the middle part of the disk, the
magnetic field is two orders of magnitude smaller than in the
corona, as predicted by our analytical solution in the disk, where
the zeroth-order field vanishes, $\vec{\mathrm B}_0=0$. 
The components are singled out in Fig.~\ref{btotcomp2} in Appendix B. Near the boundaries, our analytical solution is less exact, as expected, since it is influenced by the magnetic field in the corona. {\it Bottom panel:} ratio of the gradient of magnetic pressure $\nabla B_p^2/(8\pi)$, which measures the magnetic force, to the sum of gas pressure gradient, centrifugal, and gravity forces, in logarithmic color grading. Inside the disk, and in particular in the middle part of the disk with $r\in [5,25]R_\star$, these other forces prevail over the magnetic ones.
}
\label{btotcomp}
\end{figure}
%%+++++++++++++++++++++++++++++++++++++++++++++++++++++++++++++++++++

On these grounds, of the disk solution through the second order in $\epsilon$ being independent of the magnetic field,  we expect that if we knew the magnetic
field configuration outside the disk, the magnetic field internal to the
disk could be obtained by solving the induction equation alone on a
fixed HD background of KK00 fluid solution, as
was done in \cite{naso3}, but with the external field
furnishing here the boundary conditions on the surface of the disk. Of course, in fact we cannot know the external
magnetic field without solving for the field inside the disk as well. As is well
known, the initially dipole character of the stellar field undergoes a
considerable rearrangement, leading to what is known as field inflation,
whereby the magnetic field lines are pushed out, thus affecting the
effective boundary conditions for the induction equation solution in the
disk.

%%%%%%%%%%%%%%%%%%%%%%%%%%%%%%%%%%%%%%%%%%%%%%%%%%%%%%%%%%%%%%%%%%%
\section{Results from numerical simulations compared with analytical results}
\label{numcomp}
%%%%%%%%%%%%%%%%%%%%%%%%%%%%%%%%%%%%%%%%%%%%%%%%%%%%%%%%%%%%%%%%%%%%

%+++++++++++++++++++++++++++++++++++++++++++++++++++++++++++++++++++
\begin{figure*}
\includegraphics[scale=1,width=\columnwidth]{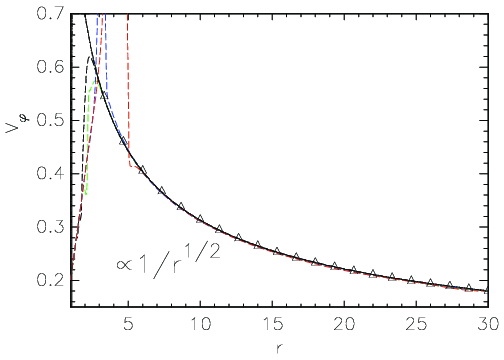}
\centering
\llap{\shortstack{\includegraphics[scale=1.,width=0.56\columnwidth]{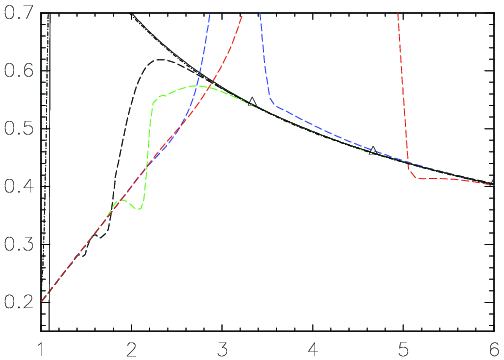}\\
\rule{0ex}{0.265\columnwidth}}
\rule{0.04\columnwidth}{0ex}}
\includegraphics[width=\columnwidth]{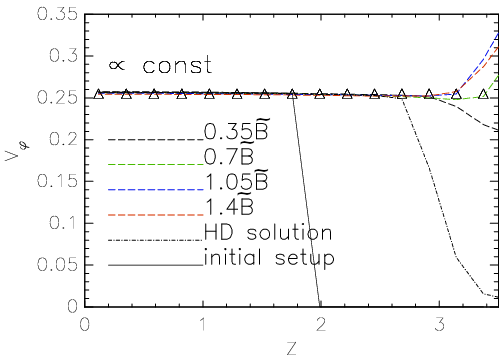} 
\caption{Azimuthal velocity of thin disks from our simulations with
$\Omega=0.2\,\Omega_{\rm ms}$. As expected from theory, it closely
follows the Keplerian curve and is uniform on cylinders
(with corrections of order $z^{2}/r^{2}$). The disk-corona interface is
at $z\approx2$. This and later figures exhibit values in the
initial set-up ({\sl thin black solid line}), and the quasi-stationary solutions
in the numerical simulations in the HD ({\sl dot-dashed black line}), and the
MHD ({\sl long-dashed line}) cases. {\sl Black, green, blue,} and {\sl red} colors
correspond to stellar magnetic field
strength ({\sl 0.35, 0.7, 1.05, 1.4}) $\tilde B$, respectively (c.f. the final paragraph of Appendix B). The closest match (in normalization) of the analytic profile to
the $0.7\tilde B$ case is depicted with {\it triangle} symbols. {\it Left panel:} radial dependence along the midplane,
just above $\theta$=$90^\circ$. Insert shows a zoom into the vicinity of the
stellar surface, where the inner boundary condition affects the solution
so that it is different than in the purely HD case. {\it Right panel:} the
profiles along a vertical line at r=15$R_\star$. The thin black solid line,
sloping down at $z\approx2$ in the right panel, indicates $v_\varphi$ in the
initial set-up. The legend describes the line styles also for the following three figures.
}
\label{fig:comparephi}
\end{figure*}
%%+++++++++++++++++++++++++++++++++++++++++++++++++++++++++++++++++++

%+++++++++++++++++++++++++++++++++++++++++++++++++++++++++++++++++++
\begin{figure*}
\centering
\includegraphics[width=\columnwidth]{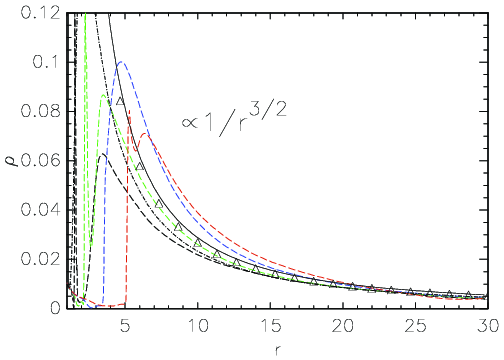}
\includegraphics[width=\columnwidth]{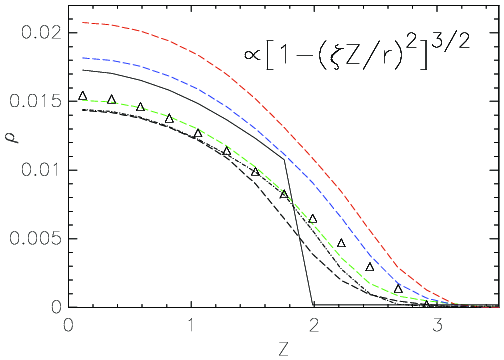}
\caption{Comparison of the matter density in the MHD simulation with analytic (triangles) and HD (dot-dashed) profiles. {\it Left panel:} radial dependence along the midplane, just above
$\theta$=$90^\circ$. {\it  Right panel:} the profiles along the vertical
line at r=15$R_\star$. The legend is given in
Fig.~\ref{fig:comparephi}.}
\label{fig:comparerho1}
\end{figure*}
%+++++++++++++++++++++++++++++++++++++++++++++++++++++++++++++++++++

%++++++++++++++++++++++++++++++++++++++++++++++++++++++++++++++++++++++
\begin{figure*}
\centering
\includegraphics[width=\columnwidth]{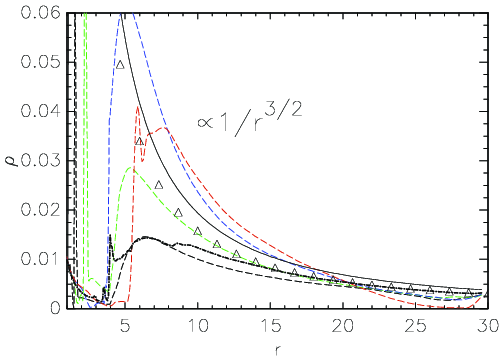}
\includegraphics[width=\columnwidth]{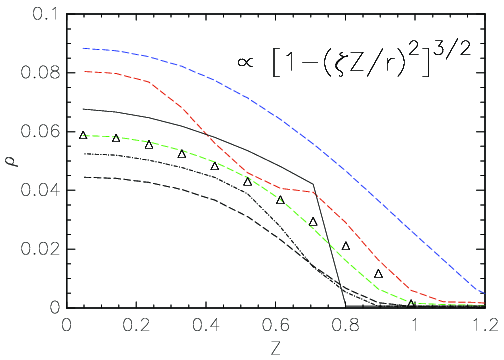}
\caption{Comparison of the matter density in the initial set-up (solid
line) with the analytic profiles (triangles) and the quasi-stationary solutions in numerical simulations
in the HD (dot-dashed line) and the MHD cases closer to the disk surface
and the star. The solution is more noisy than in the previous figure.
{\it Left panel:} density variation along
the disk surface at $\theta$=$83^\circ$. {\it Right panel:} density variation along a vertical
line at $r$=$6R_\star$; here, the trend of density increasing with the magnetic field is not maintained for the strongest fields, the red line being lower than the blue line. The meaning of the
lines is the same as in Fig.~\ref{fig:comparephi}.}
\label{fig:rhocompare2}
\end{figure*}
%
%+++++++++++++++++++++++++++++++++++++++++++++++++++++++++++++
\begin{figure*}
\centering
\includegraphics[width=\columnwidth]{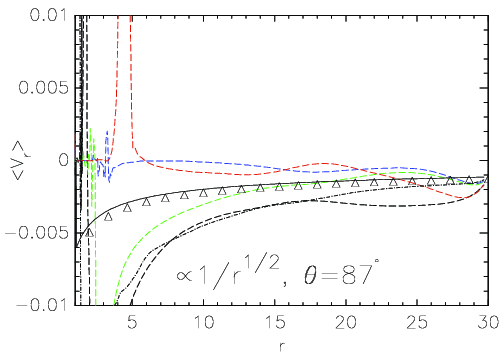}
\includegraphics[width=\columnwidth]{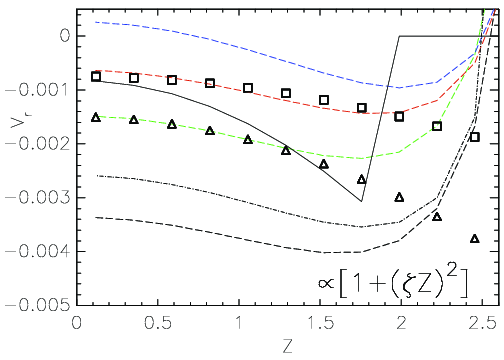}
\includegraphics[width=\columnwidth]{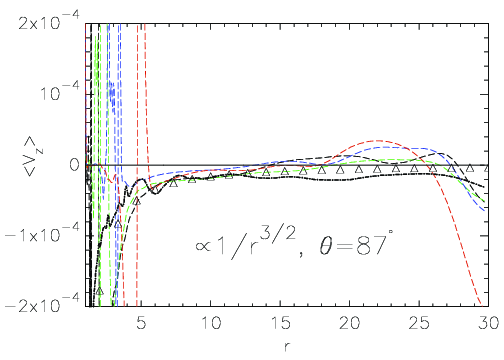}
\includegraphics[width=\columnwidth]{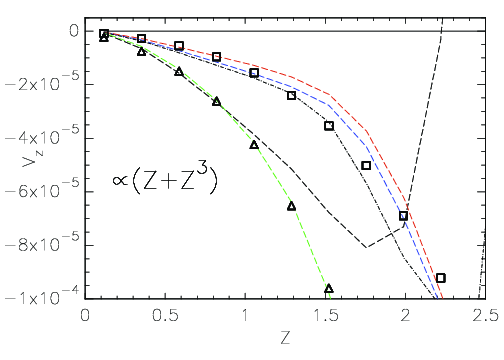} 
\caption{Comparison of the $r$ and $z$ velocity components in the
initial set-up (thin solid line) with the quasi-stationary solutions in
the numerical simulations in the HD (dot-dashed line) and the MHD
(long-dashed line) cases, with $\Omega=0.2\Omega_\mathrm{ms}$. The
meaning of the lines is the same as in Fig.~\ref{fig:comparephi}.
{\it Left panels:} Values averaged over 10 stellar rotations, to
smooth out the oscillations. {\it Right panels:} Values
at $r=15R_\star$. In the $z$-direction, an additional match
is shown with squares in the cases with larger magnetic field. It has
the same (analytically derived) functional dependence as shown with triangles for the smaller
magnetic field, only the proportionality coefficient is different.
}
\label{fig:comparerz}
\end{figure*}
%+++++++++++++++++++++++++++++++++++++++++++++++++++++++++++++
%
In our numerical simulations, which are presented in detail in \cite{cem19} and
briefly reviewed in Appendix B here, we consider a part of the star-disk
system close to the central object, with the physical domain reaching only
into the region of the disk where the Ohmic resistivity is the largest
contributor to the dissipation in the induction equation. At the outer
boundary we feed the disk with matter at a constant rate to allow the
solution to reach a quasi-stationary state, as illustrated in Appendix B.

Starting from the analytical KK00 solution as an initial condition
in the HD simulation, we obtain a numerical solution. Our computational
domain reaches into the middle disk region, where the resistivity adds
to the viscosity as a dissipation mechanism. This could make some of
the assumptions from the purely HD disk implausible---with the help of
numerical simulations we check whether or not this is true.

Our numerical solutions settle to a quasi-stationary state. A capture in our HD solution after 100 stellar rotations is shown in
Fig.~\ref{fig:hdsol} in the Appendix. In this case, the accretion onto
the star proceeds through the disk connected to the stellar equator. The
mass flux onto the star and into the polar-origin wind during the simulation
are shown in Fig.~\ref{fig:maccmominthd}.

The solution in the magnetic case is shown in Fig.~\ref{fig:diskreg}.
When the stellar dipole field is large enough, an accretion column is
formed from the inner disk rim onto the stellar surface. The matter is
lifted above the disk equatorial plane, following the magnetic field
lines. The mass flux onto the star and into the wind is shown in
Fig.~\ref{fig:maccmomint1}.

The quasi-stationary solution does not change much during the short interval
of a few tens of stellar rotations which we capture in our simulations.
The magnetic field and the accretion rate of the observed stars are
practically constant during such an interval, so that it is sensible to compare the simulations with our time-independent
analytical solutions, which---as it turned out---are a good representation of the simulation results. Comparing the quasi-stationary solutions in both the HD and MHD
solutions, to the analytical solution, we find that the MHD and HD solutions follow the analytic solutions in the
functional dependence of density and the components of velocity, only the proportionality constants change.

Before turning to a detailed comparison of the functional form
$F(r,z)$ of our numerical results with the predictions of asymptotic
expansion, we verify that the magnetic field is compatible with the
basic result of the analytical solution that the zeroth order magnetic
field vanishes, $\vec{\mathrm B}_0=0$ (``Overview'' preceding Eq.~\ref{overview},
Appendix A). Above the disk, and in the region
between the star and the disk, the field can be strong, but the values
of the magnetic field threading the {\sl thin} disk are much lower.
As seen in Fig.~\ref{btotcomp}, in the middle part of the disk the
field is two orders of magnitude smaller (dark blue region) than in
the coronal region with the stellar wind above the disk. As can be
seen in the bottom panel of the same figure, the magnetic force (gradient of the magnetic pressure) is negligible inside the disk in comparison to the other forces. Figs.~\ref{btotcomp2} and \ref{plasbeta} in Appendix B present the $B_r$, $B_\phi$, and $B_z$ components, and the plasma $\beta$, respectively.

In \cite{cem19} we varied the stellar rotation rate, magnetic field strength,
and resistivity in the disk and examined how the changes in results depend
on those parameters. Here we focus on the comparison of the results with
the conditions obtained from the analytical work. We then amend the lacking
information in the analytical solution, merging our results from the
analytical and numerical work, to obtain expressions which could be used as
a model of a magnetized thin accretion disk\footnote{In \cite{cpk19} is presented one initial example of our search for numerical expressions from our simulation results, matching the analytical solutions.}

We output the results along the disk height at two positions in the disk:
in the middle of the radial domain, at $15R_\star$,
which is in the outer region of our disk\footnote{The distance $r_m$
where the viscous torque vanishes defines a natural length scale
$r_+=\Omega_m^2r_m^4/(GM_\star)$, as shown in KK00. The outer region of the
disk is at a much larger radius.} and in the inner region of the disk,
closer to the star, at $6R_\star$, about twice the corotation radius
$r_{\rm cor}=2.9R_\star$. We derive two sets of expressions along the
vertical direction from those results, one at each distance from the
star. Along the spherical radial direction, we output the results in
the disk along a line near to the disk equator, and also along a line
near to the disk surface. For each physical quantity, we verify if
there is a unique solution throughout the disk.

Some quantities in the magnetic disk coincide with the HD solution
more than others: the best match between our HD and MHD simulations is obtained for the vertical and radial
profiles of the azimuthal velocity component (Fig.~\ref{fig:comparephi}). We note
(right panel) that the azimuthal velocity solution is uniform over a much
larger range of $z$ than in the assumed initial conditions, in which the
corona was taken to be static for $z>2$ at $r=15R_\star$, i.e., above the
initial height of the disk. Clearly, uniform rotation on vertical cylindrical surfaces is a property of the physical solution, and not an artefact of the initial conditions. At the same time the magnetic field lines are not vertical (left panel of Fig.~\ref{fig:diskreg}). These two facts taken together show that the magnetic field configuration inside the disk is controlled by the flow, and not vice versa, a conclusion which is in agreement with our analytic considerations.

In the magnetic case, the boundary conditions are different from the HD case firstly in
that $v_\varphi/r$ has to match the rotation rate at the stellar surface,
$\Omega_\star=0.2$, and secondly that now the disk surface boundary
condition at $z=z_\mathrm{d}(r)$ is not that of vacuum. The former is responsible
for\footnote{We note that $r$ is in units of $R_\star=1$.}
$v_\varphi\rightarrow 0.2\cos\theta_\mathrm{cap}$, as
$r\rightarrow \cos\theta_\mathrm{cap}$ (where $\theta_\mathrm{cap}$ is
the polar angle of the cap onto which the fluid is channeled by the
magnetic field) and the overshoot of $v_\varphi$ above the Keplerian
curve (clearly visible for the red dashed curve in Fig.~\ref{fig:comparephi}),
the latter for a different normalization of density and a different effective
height of the disk.

With $\rho_{\mathrm c0}$ typically $\sim0.01$, 
%the ratio between the initial corona and disk equatorial density, both 
the height of the disk can be
enhanced by several tens of percent (for $h'\sim 0.1$), as is the
equatorial density through the $\zeta^{-2}$ pre-factor of Eq.~(\ref{cskvadr}). This is
illustrated in Fig.~\ref{fig:comparerho1} for the midplane density,
and the vertical profile of the density at $r=15R_\star$. Because
of the difference in the influence of the boundary conditions in the
magnetic case, both the radial and vertical profiles closer to the star,
at $r=6R_\star$ and at the disk surface are more noisy
(Fig.~\ref{fig:rhocompare2}).

In agreement with our asymptotic expansion results, we find that the magnetic field does not strongly affect the flow inside the disk in the simulation. The agreement is seen by the good match between the analytic profiles (triangles) and the MHD results (lines) in Figs.~\ref{fig:comparephi},\,\ref{fig:comparerho1},\,\ref{fig:rhocompare2},\,\ref{fig:comparerz}. Accordingly,
our simulated field
configuration near the middle of the disk, around $r=15R_\star$
in Fig.~\ref{fig:diskreg}, is similar to the corresponding figure
in \cite{naso2} (Fig.~3 there), even though they solved the induction equation on a background of constant flow of the thin HD disk solution, while our magnetic field configuration and the velocity field were solved for simultaneously (self-consistently).

If the magnetic field is influencing the solutions only quantitatively,
but not qualitatively, the functional form of the fluid variables is
expected to be as in KK00. Only the density will change, because of the
inclusion of the corona-disk interface boundary conditions. The proper
form for the density equation is given by the form of Eq.~(\ref{cskvadr}),
leading to the substitution $h\rightarrow\bar h$, where $\bar h$, given by Eq.~(\ref{effh}), is a slowly varying function of
radius $r$, somewhat dependent on the magnetic field.

The analytic solutions are then:
\beqa
\rho(r,z)&\propto&\frac{1}{r^{3/2}}[\bar h^2-z^2]^{3/2},\label{compeqsa}\\  
v_r(r,z)&\propto&\frac{1}{r^{1/2}}[1+(\zeta z)^2],\label{compeqsb}  \\ 
v_z(r,z)&\approx&\frac{z}{r}v_r(r,z),\label{compeqsc} \\
v_{\varphi}(r,z)&\propto&\frac{1}{\sqrt{r}}.\ \label{compeqsd}
\eeqa

To investigate how much the magnetic solutions depart from the HD ones,
and from the KK00 analytical solution, we directly compare the density and
velocity profiles (Figs.~\ref{fig:comparephi},\,\ref{fig:comparerho1},\,\ref{fig:rhocompare2},\,\ref{fig:comparerz}). Since the analytic solutions
are obtained in the cylindrical coordinates, which are more convenient to
plot, we project our results from the simulations in spherical coordinates
to the cylindrical coordinates. In all the cases, as in
Fig.~\ref{fig:comparerho1}, we show (with triangle-symbol curves) the closest matches to the case with $B_\star=0.70\,$ $\tilde{B}$, which is the medium value in
our set of simulations.
These are
not formal best fits, but the functions in  Eqs.~(\ref{compeqsa}-\ref{compeqsd}) following from the analytic solution, with a normalization chosen to match the quasi-stationary
solution.
Inspection of the numerical solutions at $r=15R_\star$ near the disk midplane,
shown in Figs.~\ref{fig:comparephi}-\ref{fig:comparerz}, gives the following
results:
\begin{itemize}
\item[--] The radial dependence of density is practically the same in all
simulations, and is equal to the non-magnetic case.

\item[--] Vertical velocity component $v_{\mathrm z}$ also matches the
corresponding non-magnetic solution very well, while the radial component
$v_{\mathrm r}$ is more noisy.

\item[--] The vertical dependence of density shows a trend: with larger
magnetic field, the density is larger, so that it increases by up to about
25\% with the doubling of the magnetic field strength. Higher above the disk
midplane, the trend becomes more noisy.

\item[--] Both the radial and vertical dependence of the azimuthal velocity
$v_\varphi$ is excellently fitted by Eq.~(\ref{compeqsd}), except at the
very innermost parts of the disk, where the influence of the inner boundary
condition is felt. In particular, there is no $z$ dependence (rotation is
uniform on cylinders), so that our solution behaves as predicted by the
Taylor-Proudman theorem.

\item[--] The $z$ dependence of the poloidal velocity components is fairly
well described by Eqs.~(\ref{compeqsb}) and (\ref{compeqsc}). There is also
a clear trend in those components (as long as the geometry of the solutions stays the same; we describe solutions with different geometry directly below):
with larger magnetic field, values are increasing. Higher above the disk
midplane, the trend becomes more noisy.

\end{itemize}

%++++++++++++++++++++++++++++++++++++++++++++++++++++++++++++++++++++++
\begin{figure}
\includegraphics[width=\columnwidth]{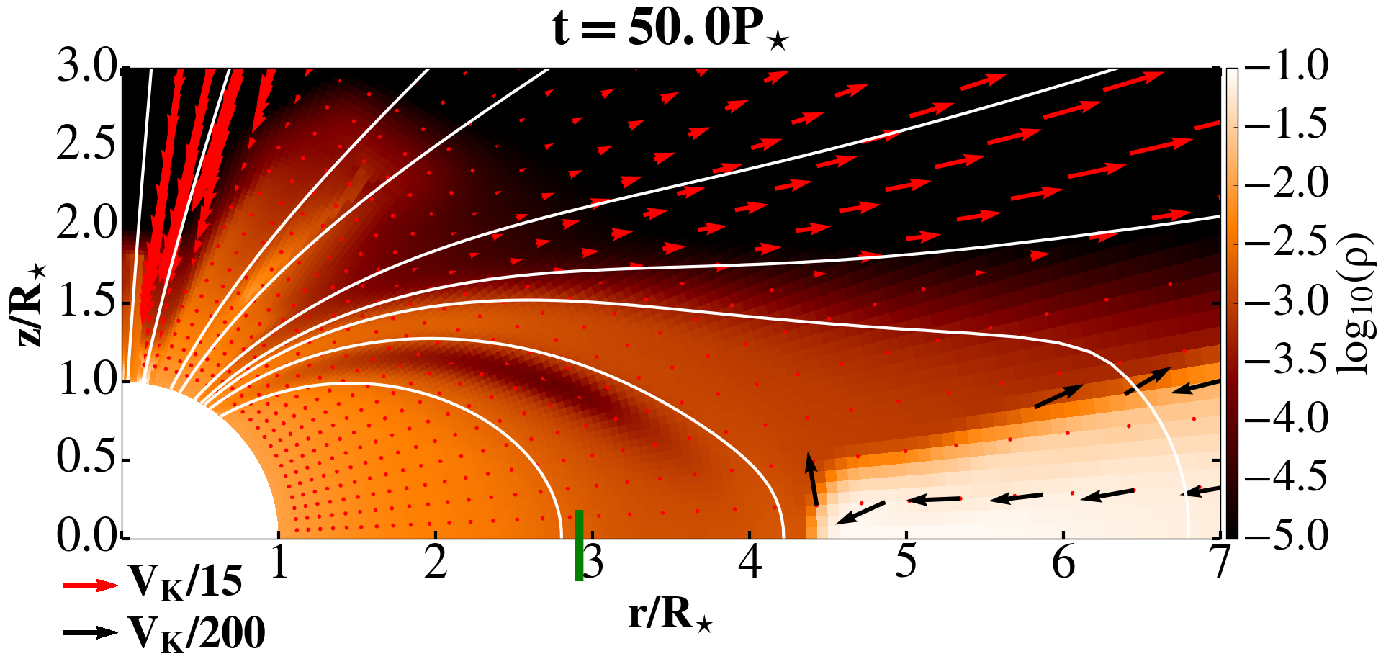}
\caption{Zoomed snapshot of the matter density in our simulation with
large magnetic field, $B_\star=1.4\tilde{B}$. Meaning of the color and lines
is the same as in Fig.~\ref{fig:diskreg}. The geometry of the solution
changes so that the magnetic pressure pushes the disk away from the
star, well beyond corotation radius. There is no accretion funnel from
the disk onto the star any more.
}
\label{fig:nocol}
\end{figure}
%++++++++++++++++++++++++++++++++++++++++++++++++++++++++++++++++++++++

In the solutions with larger magnetic field, the disk is pushed away, as
shown in Fig.~\ref{fig:nocol}, and there is no accretion column: the geometry
of the solution changed. The radial dependence departs from the predictions of
 Eqs.~(\ref{compeqsc}) and (\ref{compeqsd}), so that the $v_z(r)$ plot (at a fixed value of $\theta$) becomes
 very noisy, as do also the $v_r(r)$ curves for the two strongest magnetic field
 values (Fig.~\ref{fig:comparerz}).
 
 For the two lower $B$-field values, the $v_r(r)$ curves approximately
 track that of the HD simulation.
In agreement with our analytic analysis that was discussed in \S~\ref{asympt},
there is very little difference between the HD and these two MHD cases. Those three latter curves 
(black dot-dashed and black and green dashed lines, respectively, in
Fig.~\ref{fig:comparerz}) are significantly steeper
than the $v_r\propto 1/\sqrt{r}$ dependence of the thin disk without corona.
It would seem that the radial dependence of $v_r$ is a result of the coronal
pressure boundary condition on the disk surface. For a uniform and constant
mass accretion rate, $\dot M=\,$const,  one has $v_r\propto\dot M/(r\Sigma )$,
 where $\Sigma=\int \rho \,dz=h\rho + {\cal O}\left(h^3/r^3\right)$
 is the surface density of the disk in the upper half-plane.
 With $\rho\propto r^{-3/2}$, this gives close to the equatorial plane,
 up to terms ${\cal O}\left(h^3/r^3\right)$,
 \beq
 v_r\propto\left(\frac{r}{h}\right)r^{-1/2}.
 \eeq
 In our simulation, instead of $h\propto r$, we have an effective height 
 $\bar h$ (Eq.~\ref{effh}), which grows more rapidly than $r$, as can be
verified by the concave shape of the disk in Figs.~\ref{fig:diskreg} and
\ref{fig:hdsol}, and therefore we expect the radial velocity to also vary
more rapidly than $r^{-1/2}$.

%%%%%%%%%%%%%%%%%%%%%%%%%%%%%%%%%%%%%%%%%%%%%%%%%%%%%%%%%%%%%%%%%%%%%%%%
\section{Torques on the star}
%%%%%%%%%%%%%%%%%%%%%%%%%%%%%%%%%%%%%%%%%%%%%%%%%%%%%%%%%%%%%%%%%%%%%%%%
\label{torq}
%+++++++++++++++++++++++++++++++++++++++++++++++++++++++++++++++++++
\begin{figure*}
  \includegraphics[width=\columnwidth]{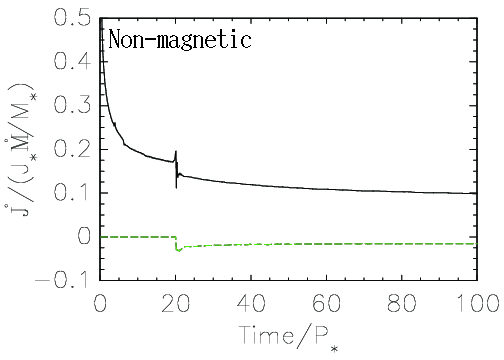}
  \includegraphics[width=\columnwidth]{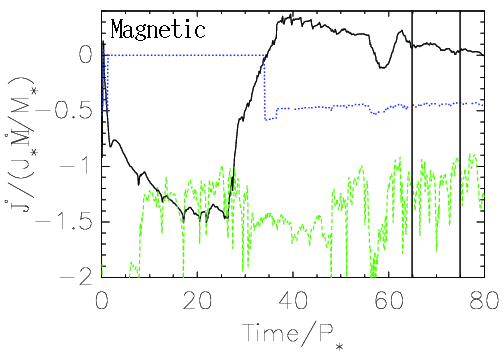}
  \caption{{\it Left panel:} accretion (``material'') torque exerted
  on the non-magnetic star by the matter accreted from the disk (solid
  line), and the angular momentum loss rate in the wind (dashed line),
  in units of the stellar angular momentum multiplied by the accretion
  timescale. {\it Right panel:} torques  exerted onto a magnetized
  star in the different components: flow in the wind (dashed green line),
  the contribution from the parts of the disk beyond $R_{\mathrm{cor}}$ 
  (dotted blue line), and within $R_{\mathrm{cor}}$ (solid black line).
The torque was measured in units of
$\dot{J}_0$=$\rho_{\rm d0}R_\star^3v_{\rm K\star}^2$. The sign convention
is such that a positive torque spins up the star. Vertical solid lines mark
the time-interval in which we consider the solution as quasi-stationary. For
comparisons of different cases, we average the physical quantities over such
intervals.
}
\label{fig:torque}
\end{figure*}
%+++++++++++++++++++++++++++++++++++++++++++++++++++++++++++++++++++

Our simulation allows us to compute the torques on the star resulting
from the disk-magnetosphere interaction including the effects of field
inflation outside the disk, and the resulting change in the magnetic
field threading the disk.

We begin with the HD case, where neither the star nor the disk has any
appreciable magnetic field (Fig.~\ref{fig:torque}). The torques are
normalized to $J_\star\dot M/ M_\star$, with
$J_\star\equiv M_\star R_\star^2\Omega_\star$ the stellar moment of
inertia computed in the shell approximation for the angular momentum of
the star. As can be seen from the figure, the torques tend to a stationary
value. The spin-up (or spin-down) timescale following from that value is on
the order of ten accretion times $ M_\star/\dot M$.

In the magnetized case, the time variation in the simulation of three
significant components of the torque on the star are illustrated in the
right panel in Fig.~\ref{fig:torque}: the torque mediated by the open field
lines, equal to the negative of the rate at which angular momentum is carried
away by the wind, magnetic torque mediated by the closed field lines terminating
on the disk beyond the co-rotation radius, and the torque exerted on the
star by the disk inside the co-rotation radius. The latter component includes
the difference in the rate of angular momentum carried by the accreting fluid
as it leaves the disk and its value as it settles on the stellar surface
(``material'' torque), as well as the magnetic interactions between the star
and the part of the disk that is inside the co-rotation radius. We have
checked that the fluid settling on the star exerts a negligible torque (as its
angular frequency is very close to that of the fairly slowly rotating star).

As in the non-magnetic case, the torques reach quasi-stationary values, with
the exception of the wind torques, which are more noisy because of the ongoing
reconnection events in the disk corona and occasional releases of energy as a result of instabilities near the stellar surface or along the outer disk
boundary. The result we obtain is that, in the magnetic case, the magnitude of
the net torque in our simulation is enhanced by an order of magnitude with
respect to the HD case accretion torque. The timescale for change of stellar
angular momentum is comparable to the accretion timescale,
$|J_\star/\dot J|\sim M_\star/\dot M$.

%%%%%%%%%%%%%%%%%%%%%%%%%%%%%%%%%%%%%%%%%%%%%%%%%%%%%%%%%%%%%%%%%%%%%%%%
\section{Discussion and conclusions}\label{concl}
%%%%%%%%%%%%%%%%%%%%%%%%%%%%%%%%%%%%%%%%%%%%%%%%%%%%%%%%%%%%%%%%%%%%%%%%
We combine the analytical conditions obtained from the asymptotic approximation
with the results from numerical simulations, to discuss solutions in resistive
MHD for a thin accretion disk threaded by a magnetic field.

Interaction of the stellar magnetosphere with the accretion disk is of particular interest in models for torque-related changes in the spin of X-ray pulsars. The position of the accretion disk inner edge is consistent between different models within a factor of two, but the discussion is still open on the nature of magnetic interaction governing the stellar rotation. \cite{naso3} questioned the assumption that the contribution to the pulsar spin-up depends on the position of the accretion column inside of the disk truncation radius, but without a fully dynamical treatment only the induction equation was considered. Here we make a step closer to a full dynamical solution. 

In the magnetic case, equations in the disk cannot be solved without
knowing the solution in the corona above the disk and between the disk and
the stellar surface. This is because of the connectivity of the magnetic field
in the disk to the stellar surface. The solution is further complicated
with the magnetic reconnection taking place in the corona. However,
the numerical simulation takes account of the disk-magnetospheric
interaction, so we solve self-consistently for the magnetic field
configuration and for the disk fluid variables. From the simulations
we obtain quasi-stationary solutions for a magnetic geometrically thin
disk.

We also performed analytic calculation, solely within the disk, allowing
a non-zero pressure at the disk-corona and magnetosphere interface.
By performing an analysis of the relevant equations written as a power
series in the small parameter $\epsilon=\tilde H/\tilde R$ (the
characteristic dimensionless disk thickness), we were able to show that
the fluid variables satisfy the same equation of motion in the MHD and
HD cases. The latter equations have been derived and solved for a
polytropic disk in vacuum by KK00. Thus we compare our simulated
solutions to the KK00 functional forms of the fluid variables. With
the exception of radial velocity, which seems to show the greatest
effect of non-vacuum boundary conditions (non-vanishing pressure at
the disk-corona interface), the simulated solutions match the analytic
expressions rather well.

We introduced the induction equation for the first time to the asymptotic
expansion analysis. To check if our removal of the dissipative, heating
terms, from the energy equation is consistent with other equations, we
add the energy equation in the asymptotic expansion, also for the first time. As it turns out, the energy equation does not yield any new constraints on the asymptotic solution in the lowest orders. We find that the solutions obtained from the energy equation to lowest orders only give conditions on the magnetic field that already follow from the induction equation and the $\nabla\cdot B=0$ condition, which jointly imply $B_{i0}=0$.

Far from the inner boundary, say for $r>10R_\star$ in our simulations,
we find the following:

$\bullet$ Solution for the radial dependence of density is practically
the same in non-magnetic and magnetic cases.

$\bullet$ The vertical dependence of density shows a trend: with larger
magnetic field, the density is larger, so that it increases for up to
about 25\% with doubling of the magnetic field strength. Higher above
the disk midplane, the trend becomes more noisy.

$\bullet$ The azimuthal velocity
$v_\varphi$ is independent of the magnetic field, with both the radial and vertical dependence excellently fitted
by Eq.~(\ref{compeqsd}), velocity inversely proportional to $\sqrt{r}$, with no dependence on $z$.

$\bullet$ Vertical velocity component, $v_z$, matches the corresponding
non-magnetic solution very well. The radial component is more noisy, especially
for high values of $B$, when the velocities are suppressed.

$\bullet$ The vertical ($z$) dependence of the poloidal velocity components is
fairly well described by Eqs.~(\ref{compeqsb}) with a fitted constant $\zeta$ [c.f. the prediction of Eq.~(\ref{zeta}) and (\ref{compeqsc})]. There is a trend in the velocity
components for those solutions which show the accretion column: the values increase for larger magnetic
field. Higher above the disk midplane, the trend
becomes more noisy.

$\bullet$ In the solutions with larger magnetic field, the disk is
pushed away and there is no accretion column. Then the $v_r$ and $v_z$ radial
dependence becomes very noisy.

$\bullet$ As it settles on the star, the fluid in our simulations exerts a
negligible torque, since by then it almost co-rotates with the  rotating star.
The ``material'' accretion torque is mediated by the magnetic fields, as the fluid flows through the accretion column.

$\bullet$ The net spin-up torque in the magnetic case is an order of magnitude
larger than in the HD case. The timescale for stellar angular change
momentum is comparable to the accretion timescale.

In this paper, we compared the analytical solutions with numerical
solutions in the cases with stellar rotation equal to 20\% of the
equatorial mass-shedding velocity. As shown in \cite{cem19}, solutions
with smaller stellar rotation rates follow similar trends, so that our
conclusions should extend to such cases. 

\par\noindent We end with some caveats: Our study here is limited to the
values of the free parameters of
viscosity and resistivity $\alpha_{\rm v}=\alpha_{\rm m}=1$. We leave
 for a separate study an investigation of the solutions in other cases,
in particular the case with a smaller viscosity parameter
$\alpha_{\rm v}<0.685$, which shows a backflow region in the disk close
to the disk equatorial plane \citep{MishraR22}. We also leave for a separate study the
cases with faster rotating stars, as they often exhibit axial jets and
conical outflows, changing the geometry of the solutions. 

We performed simulations in a quadrant of the meridional plane, enforcing
at $\theta=\mathrm{\pi}/2$ (i.e., $z=0$) boundary conditions
consistent with equatorial plane reflection symmetry. The analytic
calculation did not assume such a symmetry. However, some conclusions of
the asymptotic expansion analysis (such as the vanishing of $B_{z0}$)
relied on certain (finiteness) conditions applying in the lower half-plane,
%\citep[just as in the HD asymptotic expansion, analyzed in ][]{Borumand1997},
which is not present in our simulation. It remains to check the difference
between solutions for simulations performed in the upper half and the
full meridional plane.

\section*{Acknowledgements}
Work at CAMK is funded by the Polish NCN grant 2019/33/B/ST9/01564. M\v{C} acknowledges the Czech Science Foundation (GA\v{C}R) grant No.~21-06825X and the support by the International Space Science Institute (ISSI) in Bern, which hosted the International Team project \#495 (Feeding the spinning top) with its inspiring discussions. M\v{C} developed the setup for star-disk simulations while in CEA, Saclay, under the ANR Toupies grant, and his collaboration with Croatian STARDUST
project through HRZZ grant IP-2014-09-8656 is also acknowledged. During part of the text finalization M\v{C} was supported by CAS President’s International Fellowship for Visiting Scientists (grant No. 2020VMC0002). VP's work was partly funded by the Polish National Science Centre grant
2015/18/E/ST9/00580. We thank IDRIS (Turing cluster) in Orsay, France, ASIAA/TIARA (PL and XL clusters) in Taipei, Taiwan, and NCAC (PSK and CHUCK clusters) in Warsaw, Poland, for access to Linux computer clusters used for the high-performance computations. The {\sc pluto} team is thanked for the possibility to use the code. We thank CAMK erstwhile Ph.D. student D. A. Bollimpalli and summer students F. Bartoli\'{c} and C. Turski for developing the initial version of our Python scripts for visualization. We thank J. Hor\'{a}k for discussions of analytic solutions.

%%%%%%%%%%%%%%%%% APPENDIX %%%%%%%%%%%%%%%%%%%%%

\begin{appendix}
\section{Asymptotic expansion equations for a thin accretion disk}
An example of the asymptotic approximation method is worked-out in detail
in an example with the continuity equation, in which we derive all the
terms to the second order. The remaining equations are derived by following
the same method. We present second order equations of the set from
\S~\ref{asympt}. Unlike in the HD case (KK00), in general these cannot be
solved without additional assumptions and/or boundary conditions.

In the hydrodynamic solution one can assume that the disk density
decreases smoothly to zero towards the disk surface, which greatly
simplifies the solution. In the magnetic case, the disk solution cannot
be given without inclusion of the stellar corona, because of a magnetic
connection with the star and a corona above the disk. To obtain a
solution for the magnetic field penetrating the disk, we have to include
the disk-corona boundary condition, which is unknown. Because of this, we
can obtain only the most general conditions for the disk magnetic field from
the equations. Full information about the magnetic field solution inside
the disk can be obtained from our numerical simulations.

The normalization is defined with the following equations:
$\epsilon=\tilde{c}_{\mathrm s}/(\tilde{R}\tilde{\Omega})=\tilde{H}/\tilde{R}\ll 1$,
so that $\tilde{c}_{\mathrm s}=\epsilon\tilde{R}\tilde{\Omega}$,
and then
$c_s'=c_s/\tilde{c}_{\mathrm s}=c_s/(\epsilon\tilde{R}\tilde{\Omega})$.
Tildes denote characteristic values of the variables,
and primes the scaled variables. Here, $\tilde{R}$ is the cylindrical radius, so $\tilde{v}_\phi\equiv\tilde{R}\tilde{\Omega}$ is the characteristic azimuthal velocity.
Further,
$\Omega'=\Omega/\tilde{\Omega}$,
$\tilde{\Omega}=\Omega_\mathrm{ms}=\sqrt{GM_\star/\tilde{R}_\star^3}$,
$r'=r/\tilde{R}$, $z'=z/(\epsilon\tilde{R})$,
$\varv_r'=\varv_r/\tilde{c}_{\mathrm s}=\varv_r/(\epsilon\tilde{R}\tilde{\Omega})$,
$\varv_z'=\varv_z/\tilde{c}_{\mathrm s}=\varv_z/(\epsilon\tilde{R}\tilde{\Omega})$,
$\varv_\varphi'=
% not \varv_\varphi/\tilde{c}_{\mathrm s}=
\varv_\varphi/(\tilde{R}\tilde{\Omega})$.
The magnetic field we normalize with the Alfv\'{e}n speed
$\tilde\varv_{\mathrm A}=\tilde{B}/\sqrt{4\pi\tilde{\rho}}$ as a characteristic
speed, and $\rho'=\rho/\tilde{\rho}$. Then we have
$B'=B/\tilde{B}=B/({\tilde\varv}_{\mathrm A}\sqrt{4\pi\tilde{\rho}})$, and
$\tilde{B}$ is the normalization for all the magnetic field components:
$B_r'=B_r/\tilde{B}$, $B_z'=B_z/\tilde{B}$,
$B_\varphi'=B_\varphi/\tilde{B}$.

The beta plasma parameter
$\beta=P_{\mathrm {gas}}/P_{\mathrm {mag}}=8\pi P_{\mathrm {gas}}/B^2$. With
$P=P_{\mathrm {gas}}$ we can write
$c_s^2=\gamma P/\rho=\gamma \beta B^2/(8\pi\rho)=\gamma\beta \varv_{\mathrm A}^2/2$,
so that $\tilde{\varv}_{\mathrm A}^2/\tilde{c}_{\mathrm s}^2=2/(\gamma\tilde{\beta})$.

The viscosity scales with the sound speed as a characteristic velocity and
the height of the disk $H$, so that the normalization for the kinetic
viscosity is $\tilde{\nu}_{\mathrm
v}=\tilde{c}_{\mathrm s}\tilde{H}=\epsilon^2\tilde{R}^2\tilde{\Omega}$,
and then $\tilde{\eta}=\tilde{\rho}\tilde{\nu}_{\mathrm v}=
\tilde{\rho}\epsilon^2\tilde{R}^2\tilde{\Omega}$. Then
$\eta'=
\eta/\tilde{\eta}=\eta/(\tilde{\rho}\epsilon^2\tilde{R}^2\tilde{\Omega})$.
For the resistivity we choose the normalization with the Alfv\'{e}n speed as a
characteristic speed, so that
$\tilde{\eta}_{\mathrm m}=\tilde{\varv}_{\mathrm A}\tilde{H}=
\epsilon\tilde{R}\tilde{\varv}_{\mathrm A}$.
Then $\eta'_{\mathrm m}=\eta_{\mathrm m}/\tilde{\eta}_{\mathrm m}=
\eta_{\mathrm m}/(\epsilon\tilde{R}\tilde{\varv}_{\mathrm A})=
\eta_{\mathrm m}\left(
\sqrt{\gamma\tilde{\beta}/2}\right)/(\tilde{c}_{\mathrm s}\epsilon\tilde{R})
=\eta_{\mathrm m}\left(
\sqrt{\gamma\tilde{\beta}/2}\right)/(\epsilon^2\tilde{R}^2\tilde{\Omega})$.

In the asymptotic approximation, following KK00, we write all the variables in the Taylor expansion with the coefficient of expansion $\epsilon=\tilde{H}/\tilde{R}<<1$. For a variable X, we then have
$X=X_0+\epsilon X_1+\epsilon^2 X_2+\epsilon^3 X_3+\dots $, and we can compare
the terms of the same order in $\epsilon$.
We write the normalized equations of magnetic field
solenoidality, continuity,
induction, momentum, and energy density, i.e., Eqs.~(\ref{eq1})--(\ref{eq5}).
For simplicity, in some cases we use the notation
$\partial_x=\partial/\partial x$, and we drop all primes in the following
(where all the variables are scaled, so no confusion can arise).

We will be searching for stationary solutions, so that the additional
conditions are that of stationarity, $\partial/\partial t=0$,
and axial symmetry $\partial/\partial\varphi=0$. We work in the
cylindrical coordinates $(r, z, \varphi)$.
%%%%%%%%%%%%%%%%% %%%%%%%%%%%%%%%%% %%%%%%%%%%%%%%%%% 
\subsection*{Equation of continuity}
We start from the continuity equation, Eq.~(\ref{eq2}) in \S~\ref{asympt}:
\beq
\frac{\partial\rho}{\partial t}+\nabla\cdot(\rho\mathbf{v})=0.
\nonumber
\eeq
In the stationary case, when $\partial_t\rho=0$, and applying also the
axi-symmetry condition $\partial_\varphi(\rho\mathbf{v})=0$:
\begin{equation}
  \frac{1}{r}\partial_r(r\rho\varv_r)+
\partial_z(\rho\varv_z)=0.\nonumber
\end{equation}
We can write the
normalized equation, in which the terms can be written in the orders of a
small parameter $\epsilon$:
\begin{equation}
\frac{1}{\tilde{R}r'}\frac{1}{\tilde{R}}\partial_{r'}
(r'\tilde{R}\tilde{\rho}\rho'\epsilon\tilde{\Omega}\tilde{R}\varv_{r'})+
\frac{1}{\epsilon\tilde{R}}\partial_{z'}\rho\tilde{\rho}\epsilon\tilde{\Omega}\tilde{R}\varv_{z'}
=0\ .\nonumber
\end{equation}
Removing the primes, we can write:
\begin{equation}
\frac{\epsilon}{r}\partial_r(r\rho\varv_r)+\partial_z(\rho\varv_z)=0.\nonumber
\end{equation}
Writing the expansion in $\epsilon$ in each quantity, we obtain
\begin{eqnarray}
\frac{\epsilon}{r}\partial_r[r(\rho_0+\epsilon\rho_1+\epsilon^2\rho_2+\dots)(\varv_{r0}+\epsilon\varv_{r1}+\epsilon^2\rho_2+\dots)]\nonumber \\
+\partial_z[(\rho_0+\epsilon\rho_1+\epsilon^2\rho_2+\dots)(\varv_{z0}+\epsilon\varv_{z1}+\epsilon^2\varv_{z2}+\dots)]=0.\nonumber
\end{eqnarray}
From this we can write the term in the order zeroth order in $\epsilon$ as:
\subsubsection*{Order $\epsilon^{0}$:}
\begin{equation}
  \frac{\partial}{\partial z}\left(\rho_0\varv_{z0}\right) = 0 
  \Rightarrow\varv_{z0} = 0.
  \label{cont0}
\end{equation}
In a stationary solution, the disk midplane cannot be suffering a steady
vertical drift,
so we must have $\rho_0\varv_{z0}=0$ at ${z=0}$.
Since the product does not depend on  $z $,
we conclude that we must have $\varv_{z0}=0$ throughout the disk
($\rho_0\neq 0$).

\subsubsection*{Order $\epsilon^{1}$:}
In the first order in $\epsilon$ we have:
\begin{equation}
\frac{1}{r}\frac{\partial}{\partial r}\left(r\rho_0\varv_{r0}\right) + \frac{\partial}
     {\partial z}\left(\rho_0\varv_{z1}\right) = 0
     \Rightarrow\varv_{z1}=0.
\end{equation}
As we will see from the radial momentum equation,  in the first order in
$\epsilon$, Eq.~(\ref{azmom1}), we have $\varv_{r0}=0$, so that here we
have $\partial_z(\rho_0\varv_{z1})=0\Rightarrow \rho_0\varv_{z1}=
\mathrm{const}$ along $z$. Following the same argumentation as above,
we conclude that $\varv_{z1}=0$.

\subsubsection*{Order $\epsilon^{2}$:}
In the second order in $\epsilon$ we have:
\begin{equation}
\frac{1}{r}\frac{\partial}{\partial r}\left(r\rho_0\varv_{r1}\right)+
\frac{\partial}{\partial z}\left(\rho_0\varv_{z2}\right) = 0.
\end{equation}
Thus, if we find a solution for $\varv_{r1}$,
we will be able to determine the $z$ dependence of $\varv_{z2}$.

The same procedure is carried in each of the remaining equations
of \S~\ref{asympt}. In the following, we will often find that certain
quantities are functions of the radial variable alone. In such cases
we will denote a {\it generic} radial function as $f(r)$, without
implying any particular functional dependence on $r$, so that the
results $a =f(r)$, and $b =f(r)$ do {\it not} imply
$a(r,z)\equiv b(r,z)$.

\subsection*{\rm {\bf Overview}}
By parity arguments (KK00), or from a more formal derivation
\citep[Appendix A in][or Borumand 1997]{Reb09},
$\Omega_1=0$. Similarly,  $c_{\mathrm s1}$=$\rho_1$=$\varv_{r0}=0$,
for all $r,z$, with the last equation also expressly following
from Eq.~(\ref{azmom1}).
It is easy to check that all the following equations are consistent with
the zeroth order components of the magnetic field  vanishing,
$B_{i0}\equiv 0$,  $i\in\{r,z,\varphi\}$, as well as $B_{z1}\equiv 0$. 

We do not claim uniqueness of the presented solution, should the magnetic pressure turn out to be so dominant that $\mathcal O(\bar\beta)=\epsilon$. However, on the assumption that $\mathcal O(\bar\beta)\ge1$ it can be shown that $B_{i0}=0$ is the only solution satisfying reflection symmetry in the equatorial plane; the proof is outside the scope of this Appendix, but in the asymptotic expansion below we obtain conditions like $\partial_z B_{i0}=0$ or $B_{j0}\partial_z B_{i0}=0$ for all the components of magnetic field, already implying that either $B_{j0}\equiv 0$ or $B_{i0}=f(r)$. The same is obtained for $B_0$ itself, and as a result the equation for the zeroth order vertical momentum also becomes independent of zeroth order total magnetic field $B_0$. Such an outcome is also supported by the results of our numerical simulations. 

%%%%%%%%%%%%%%%%% %%%%%%%%%%%%%%%%% %%%%%%%%%%%%%%%%% 
\subsection*{Condition ${\bf \nabla\cdot}\mathbf B =0$:}
\begin{equation}
\frac{\epsilon}{r}\frac{\partial}{\partial r}
\left(r B_r\right) + \frac{\partial}{\partial z}\left(B_z\right) = 0\ .
\label{overview}
\end{equation}

\subsubsection*{Order $\epsilon^{0}$:}
\begin{equation}
  \frac{\partial B_{z0}}{\partial z} = 0\ .
\label{bzofr}
\end{equation}

\subsubsection*{Order $\epsilon^{1}$:}
\begin{equation}
  \frac{1}{r}\frac{\partial}{\partial r}\left(r B_{r0}\right)
  + \frac{\partial}{\partial z}\left(B_{z1}\right) = 0\ .
\label{divb2}
\end{equation}

%%%%%%%%%%%%%%%%% %%%%%%%%%%%%%%%%% %%%%%%%%%%%%%%%%% 
\subsection*{Energy equation:}
\begin{equation}
\begin{aligned}
\epsilon^4 \tilde{R}^2\tilde{\Omega}^2 n\rho \varv_r\frac{\partial c_s^2}
{\partial r}+\epsilon^3\tilde{R}^2\tilde{\Omega}^2 n
\rho \varv_z\frac{\partial c_s^2}{\partial z}
+\epsilon\rho\tilde{\varv}^2\varv_z \frac{\partial \varv^2/2}{\partial z}\\
+\epsilon^2\rho\tilde{\varv}^2_{\rm A}\varv_r\frac{\partial \varv_{\rm A}}{\partial r}
+\epsilon\rho\tilde{\varv}_{\rm A}^2\varv_z
\frac{\partial \varv_{\rm A}^2}{\partial z}
+\epsilon^2\rho\tilde{\varv}^2\varv_r\frac{\partial \varv/2}{\partial r}\\
+\left[\epsilon^2\rho\tilde{\Omega}^2 \tilde{R}^2\frac{\varv_r}{r^2}+\epsilon^3\rho\tilde{\Omega}^2\tilde{R}^2\frac{\varv_z z}{r^3}\right]
\left[1+\epsilon^2\left(\frac{z}{r} \right)^2\right]^{-3/2}\\
=\frac{\tilde{\varv}^2_{\mathrm A}B_r}{r}
\left(\epsilon^2 \varv_r B_r+\epsilon\Omega rB_\varphi\right)\\
+\tilde{\varv}^2_{\rm A}\frac{\partial}
{\partial r}\left(\epsilon^2 \varv_r B_r^2 +
\epsilon^2 \varv_z B_z B_r +\epsilon\Omega r B_{\varphi} B_r \right) \\
+\tilde{\varv}^2_{\rm A}\frac{\partial}{\partial z}\left( \epsilon \varv_r B_r B_z+
\epsilon \varv_z B_z^2+\Omega r B_{\varphi} B_z\right)\ .
\end{aligned}
\end{equation}

\subsubsection*{Order $\epsilon^{0}$:}
\beqa
0=\tilde{\varv}^2_{\rm A}\frac{\partial}{\partial z}(\Omega_0 r B_{\varphi 0}
B_{z0})\nonumber\\
\Rightarrow\frac{\partial }{\partial z}\left(B_{z0} B_{\varphi 0}\right)
=B_{z0}\frac{\partial B_{\varphi 0} }{\partial z}=0,
\label{eneq0}
\eeqa
with the last but one equality following from Eq.~(\ref{bzofr}),
and the last from the logical implication of nonvanishing in the zeroth order of $\Omega$ from  Eq.~(\ref{omega0}),  the zeroth order radial momentum equation.
    
\subsubsection*{Order $\epsilon^{1}$:}
\begin{equation}
\begin{aligned}
B_{\varphi 0}\left(\frac{B_{r0}}{2r}+\frac{\partial B_{r0}}{\partial r}+\frac{\partial B_{z1}}{\partial z}\right)
+B_{z0}\frac{\partial B_{\varphi 1}}{\partial z}\\ 
+B_{r0}\frac{\partial B_{\varphi 0}}{\partial r}+B_{z1}\frac{\partial B_{\varphi 0}}{\partial z}=0.
\label{eneq1}
\end{aligned}
\end{equation}

%%%%%%%%%%%%%%%%% %%%%%%%%%%%%%%%%% %%%%%%%%%%%%%%%%% 
\subsection*{Vertical induction:}
\begin{equation}
\begin{aligned}
0 = \frac{\varv_zB_r}{r}-\frac{\varv_rB_z}{r}
+B_r\frac{\partial \varv_z}{\partial r}
-\varv_r\frac{\partial B_z}{\partial r}
+\varv_z\frac{\partial B_r}{\partial r}\\
-B_z\frac{\partial \varv_r}{\partial r}
+\sqrt{\frac{2}{\gamma\tilde{\beta}}}
\Bigg(\epsilon\frac{\partial\eta_{\mathrm m}}{\partial r}\frac{\partial B_z}{\partial r}
-\frac{\partial\eta_{\mathrm m}}{\partial r}\frac{\partial B_r}{\partial z}\Bigg)\\
+\eta_{\mathrm m}\sqrt{\frac{2}{\gamma\tilde{\beta}}}
\Bigg(\frac{\epsilon}{r}\frac{\partial B_z}{\partial r}-
\frac{1}{r}\frac{\partial B_r}{\partial r}-
\frac{\partial^2 B_r}{\partial r \partial z}+
\epsilon\frac{\partial^2B_z}{\partial r^{2}}\Bigg).
\end{aligned}
\end{equation}

\subsubsection*{Order $\epsilon^{0}$:}
Because the radial and vertical components of velocity vanish in the zeroth order by Eqs.~(\ref{cont0}), (\ref{azmom1}), we are left with
\beq
\eta_{\mathrm m0}\Bigg(\frac{1}{r}\frac{\partial B_{r0}}{\partial r}+\frac{\partial^2 B_{r0}}{\partial r\partial z}\Bigg)+\frac{\partial\eta_{\mathrm m0}}{\partial r}\frac{\partial B_{r0}}{\partial z}=0,
\label{vertind0a}
\eeq
consistent with $B_{r0}=0$.

\subsubsection*{Order $\epsilon^{1}$:}
\beq
\begin{aligned}
\varv_{r1}\left(\frac{B_{z0}}{r}+\frac{\partial B_{z0}}{\partial r}\right)
+B_{z0}\frac{\varv_{r1}}{\partial r}\\  =\sqrt{\frac{2}{\gamma\tilde{\beta}}}
\Bigg[\frac{\partial\eta_{\mathrm m0}}{\partial r}\Bigg(\frac{\partial B_{z0}}{\partial r}
-\frac{\partial B_{r1}}{\partial z}\Bigg)
-\frac{\partial\eta_{\mathrm m1}}{\partial r}\frac{\partial B_{r0}}{\partial z}\\
+\frac{\eta_{\mathrm m0}}{r}\Bigg(\frac{\partial B_{z0}}{\partial r}
-\frac{\partial B_{r1}}{\partial r}\Bigg)
-\eta_{\mathrm m0}\frac{\partial^2 B_{r1}}{\partial r\partial z}\\
-\eta_{{\mathrm m1}}\Bigg(\frac{1}{r}\frac{\partial B_{r0}}{\partial r}+\frac{\partial^2B_{r0}}{\partial r\partial z}\Bigg)\Bigg].
\end{aligned}
\label{vertind1a}
\eeq
With vanishing components of $B_0$:
\begin{equation}
\frac{\partial}{\partial r}
\left(\eta_{\mathrm m0}\frac{\partial B_{r1}}{\partial z}\right)
+\frac{\eta_{\mathrm m0}}{r}\frac{\partial B_{r1}}{\partial r}=0.
\label{vertind1}
\end{equation}

%%%%%%%%%%%%%%%%% %%%%%%%%%%%%%%%%% %%%%%%%%%%%%%%%%% 
\subsection*{Radial induction:}
\begin{equation}
\begin{aligned}
0=B_z\frac{\partial\varv_r}{\partial z}+\varv_r\frac{\partial B_z}{\partial z}
-B_r\frac{\partial \varv_z}{\partial z}-\varv_z\frac{\partial B_r}{\partial z}\\
+\sqrt{\frac{2}{\gamma \tilde{\beta}}}\left(\frac{\partial\eta_{\rm m}}{\partial z}\frac{\partial B_r}{\partial z}-\epsilon \frac{\partial \eta_{\rm m}}{\partial z}\frac{\partial B_z}{\partial r}\right)\\
+\eta_{\rm m}\sqrt{\frac{2}{\gamma \tilde{\beta}}}
\left(\frac{\partial^2 B_r}{\partial z^2}-\epsilon\frac{\partial^2 B_z}{\partial r\partial z}\right).
\end{aligned}
\end{equation}

\subsubsection*{Order $\epsilon^{0}$:}
\beq
\frac{\partial\eta_{\rm m0}}{\partial z}\frac{\partial B_{r0}}{\partial z}
+\eta_{\rm m0}\frac{\partial^2 B_{r0}}{\partial z^2}=0.
\label{radinzerob0}
\eeq

\subsubsection*{Order $\epsilon^{1}$:}
\beq
\begin{aligned}
B_{z0}\frac{\partial\varv_{r1}}{\partial z}
+\sqrt{\frac{2}{\gamma\tilde{\beta}}}
\Bigg(\frac{\partial\eta_{\mathrm m0}}{\partial z}\frac{\partial B_{r1}}{\partial z}
+\frac{\partial\eta_{\mathrm m1}}{\partial z}\frac{\partial B_{r0}}{\partial z}\\
-\frac{\partial\eta_{\mathrm m0}}{\partial z}\frac{\partial B_{z0}}{\partial r}\Bigg) 
+\sqrt{\frac{2}{\gamma\tilde{\beta}}}\Bigg(\eta_{\mathrm m0}\frac{\partial^2 B_{r1}}{\partial z^2}
+\eta_{\mathrm m1}\frac{\partial^2 B_{r0}}{\partial z^2}\\
-\eta_{\mathrm m0}\frac{\partial^2 B_{z0}}{\partial r\partial z}\Bigg)=0.
\end{aligned}
\label{rradind0a}
\eeq

%%%%%%%%%%%%%%%%% %%%%%%%%%%%%%%%%% %%%%%%%%%%%%%%%%% 
\subsection*{Azimuthal induction:}
\begin{equation}
\begin{aligned}
0 = \epsilon rB_r\frac{\partial\Omega}{\partial r}
+\epsilon r\Omega\frac{\partial B_r}{\partial r}
+\epsilon\Omega B_r+rB_z\frac{\partial\Omega}{\partial z}
+r\Omega\frac{\partial B_z}{\partial z}\\
-\epsilon^2\varv_r\frac{\partial B_\varphi}{\partial r}
-\epsilon\varv_z\frac{\partial B_\varphi}{\partial z}
-\epsilon^{2} B_{\varphi} \frac{\partial \varv_r}{\partial r}
-\epsilon B_{\varphi}\frac{\partial \varv_z}{\partial z}\\
+\sqrt{\frac{2}{\gamma \tilde{\beta}}}
\left(\frac{\epsilon^3B_\varphi}{r}\frac{\partial \eta_{\mathrm m}}{\partial r}
+\epsilon^3\frac{\partial\eta_{\mathrm m}}{\partial r}
\frac{\partial B_\varphi}{\partial r}
+\epsilon\frac{\partial\eta_{\mathrm m}}{\partial z}
\frac{\partial B_\varphi}{\partial z}\right)\\
+\eta_{\mathrm m}\sqrt{\frac{2}{\gamma \tilde{\beta}}}
\left(\frac{\epsilon^3}{r}\frac{\partial B_\varphi}{\partial r}
-\frac{\epsilon^3B_\varphi}{r^2}
+\epsilon^3\frac{\partial^2B_\varphi}{\partial r^2}
+\epsilon\frac{\partial^2B_\varphi}{\partial z^2}\right).
\end{aligned}
\label{azind}
\end{equation}

\subsubsection*{Order $\epsilon^{0}$:}
\beq
r\Omega_0\frac{\partial B_{z0}}{\partial z}=0 \Rightarrow
\frac{\partial B_{z0}}{\partial z}=0,
\label{indazimu}
\eeq
in agreement with Eq.~(\ref{bzofr}).

\subsubsection*{Order $\epsilon^{1}$:}
\beq
\begin{aligned}
rB_{r0}\frac{\partial\Omega_0}{\partial r}
+r\Omega_0\frac{\partial B_{r0}}{\partial r}
+r\Omega_0\frac{\partial B_{z1}}{\partial z}\\
+\sqrt{\frac{2}{\gamma \tilde{\beta}}}
\Bigg(\frac{\partial\eta_{\mathrm m0}}{\partial z}
\frac{\partial B_{\varphi 0}}{\partial z}
+\eta_{\mathrm m0}\frac{\partial^2 B_{\varphi 0}}{\partial z^2}\Bigg)=0,
\end{aligned}
\label{azind1a}
\eeq
which, with vanishing components of $B_0$, becomes:
\beq
r\Omega_0\frac{\partial B_{z1}}{\partial z}=0.
\label{azind1}
\eeq

%%%%%%%%%%%%%%%%% %%%%%%%%%%%%%%%%% %%%%%%%%%%%%%%%%% 
\subsection*{Radial momentum:}
\begin{equation}
\begin{aligned}
\epsilon^2\varv_r\frac{\partial\varv_r}{\partial r}+\epsilon\varv_z\frac{\partial\varv_r}{\partial z}-\Omega^2r =-\frac{1}{r^2}
\left[1+\epsilon^2\left(\frac{z}{r}\right)^2\right]^{-3/2}\\
-\epsilon^2n\frac{\partial c_{\mathrm s}^2}{\partial r}+\frac{2}{\gamma\tilde{\beta}}\frac{1}{\rho}\left(\epsilon^2B_r\frac{\partial B_r}{\partial r}
+\epsilon B_z\frac{\partial B_r}{\partial z}-\epsilon^2\frac{B_\varphi^2}{r}\right)\\
-\frac{\epsilon^2}{\gamma\tilde{\beta}}\frac{1}{\rho}\frac{\partial B^2}{\partial r}+ 
\frac{\epsilon^3}{\rho r}\frac{\partial}{\partial r}\left(2\eta r\frac{\partial\varv_r}{\partial r}\right)+\frac{\epsilon}{\rho}\frac{\partial}{\partial z}\left(\eta\frac{\partial\varv_r}{\partial z}\right)\\
+\frac{\epsilon^2}{\rho}\frac{\partial}{\partial z}\left(\eta\frac{\partial\varv_z}{\partial r}\right)-\epsilon^3\frac{2\eta\varv_r}{\rho r^2}-\frac{2\epsilon^3}{3\rho}\frac{\partial}{\partial r}\Bigg[\eta\frac{1}{r}\frac{\partial}{\partial r}\Big(r\varv_r\Big)\Bigg]\\
-\frac{2}{3}\frac{\epsilon^2}{\rho}\frac{\partial}{\partial r}\left(\eta\frac{\partial\varv_z}{\partial z}\right).
\end{aligned}
\end{equation}
Here $n$ is the polytropic index. In the case of adiabatic index $\gamma=5/3$
for an ideal gas, we have $n=3/2$.
\subsubsection*{Order $\epsilon^{0}$:}
\begin{equation}
\Omega_{0} = r^{-3/2} .
\label{omega0}
\end{equation}

\subsubsection*{Order $\epsilon^{1}$:}
\begin{equation}
\begin{aligned}
  - 2r\Omega_{0}\Omega_{1} =
  \frac{2}{\gamma \tilde{\beta}}\frac{1}{\rho_{0}} B_{z0}
  \frac{\partial B_{r0}}{\partial z} +
  \frac{1}{\rho_{0}} \frac{\partial}{\partial z}
  \left(\eta_0 \frac{\partial \varv_{r0}}{\partial z}\right)\ .
\label{radmom1}
\end{aligned}
 \end{equation}

\subsubsection*{Order $\epsilon^{2}$:}
\begin{equation}
\begin{aligned}
2r\rho_0\Omega_0\Omega_2=\frac{3\rho_0}{2}\frac{z^2}{r^4}
+n\rho_0\frac{\partial c_{s0}^2}{\partial r}-\frac{\partial}{\partial z}
\left(\eta_0\frac{\partial\varv_{r1}}{\partial z}\right)\\
-\frac{2}{\gamma\tilde{\beta}}
\Bigg( B_{r0}\frac{\partial B_{r0}}{\partial r}
+B_{z0}\frac{\partial B_{r1}}{\partial z} +B_{z1}\frac{\partial B_{r0}}{\partial z}
-\frac{B_{\varphi 0}^2}{r}\Bigg)\\
+\frac{1}{\gamma\tilde{\beta}}\frac{\partial B_0^2}{\partial r}.
\label{radmom2}
\end{aligned}
\end{equation}
%%%%%%%%%%%%%%%%% %%%%%%%%%%%%%%%%% %%%%%%%%%%%%%%%%% 
\subsection*{Azimuthal momentum:}
\begin{equation}
\begin{aligned}
\epsilon \frac{\rho\varv_r}{r^2}\frac{\partial}{\partial r}\left(r^{2}\Omega\right) + 
\rho \varv_z \frac{\partial \Omega}{\partial z} = \frac{\epsilon^2}{r^3}\frac{\partial}{\partial r}
\left(r^3\eta\frac{\partial \Omega}{\partial r}\right)\\
+\frac{\partial}{\partial z}\left(\eta \frac{\partial \Omega}{\partial z}\right)
+\frac{2}{\gamma\tilde{\beta}}
\frac{1}{r}\left(\epsilon^2 B_r\frac{\partial B_{\varphi}}{\partial r} + \epsilon B_z\frac{\partial B_\varphi}{\partial z} +
\epsilon^2\frac{B_\varphi B_r}{r}\right).
\end{aligned}
\end{equation}

\subsubsection*{Order $\epsilon^{0}$:}
\begin{equation}
0 = \frac{\partial}{\partial z}\left(\eta_0\frac{\partial \Omega_{0}}{\partial z}\right),
\label{azmom}
\end{equation}
which already follows from Eq.~(\ref{omega0}).

\subsubsection*{Order $\epsilon^{1}$:}
\begin{equation}
\begin{aligned}
\frac{\rho_{0}\varv_{r0}}{r^{2}}\frac{\partial}{\partial r}\left(r^{2}\Omega_0\right) = 
\frac{\partial}{\partial z}\left(\eta_0\frac{\partial \Omega_1}{\partial z}\right) + 
\frac{2}{\gamma \tilde{\beta}}\frac{1}{r}B_{z0}\frac{\partial B_{\varphi 0}}{\partial z}\\
\Rightarrow \varv_{r0} =0\ .
\label{azmom1}
\end{aligned}
\end{equation}
Since the last term vanishes by Eq.~(\ref{eneq0}), and $\Omega_1=0$, we obtain $\varv_{r0}=0$, necessarily.
As expected, far from the inner and outer edges, radial flow in a thin disk is always subsonic.

\subsubsection*{Order $\epsilon^{2}$:}
\begin{equation}
\begin{aligned}
\frac{\rho_{0}\varv_{r1}}{r}\frac{\partial}{\partial r}\left(r^{2}\Omega_0\right)=\frac{\partial}{\partial z}\left(\eta_0\frac{\partial\Omega_2}{\partial z}\right)+\frac{1}{r^2}\frac{\partial}{\partial r}
\left(r^3\eta_0\frac{\partial \Omega_0}{\partial r}\right)
 \\
+
\frac{2}{\gamma \tilde{\beta}}\left(B_{r0}\frac{\partial B_{\varphi 0}}
{\partial r}+B_{z0}\frac{\partial B_{\varphi 1}}{\partial z}+\frac{B_{r0}B_{\varphi
0}}{r}\right).
\label{azmom2}
\end{aligned}
\end{equation}

%%%%%%%%%%%%%%%%% %%%%%%%%%%%%%%%%% %%%%%%%%%%%%%%%%% 
\subsection*{Vertical momentum:}
\begin{equation}
\begin{aligned}
\epsilon \varv_r \frac{\partial \varv_z}{\partial r} + \varv_z \frac{\partial
\varv_z}{\partial z} =-\frac{z}{r^{3}}\left[1+\epsilon^{2}\left(\frac{z}{r}\right)^{2}\right]^{-3/2}\\
-n\frac{\partial c_{\mathrm s}^2}{\partial z}+\frac{2}{\gamma\tilde{\beta}}\frac{1}{\rho}\left(\epsilon B_r\frac{\partial B_z}
{\partial r}+B_z\frac{\partial B_z}{\partial z}\right)-\frac{1}{\gamma\tilde{\beta}}
\frac{1}{\rho}\frac{\partial B^{2}}{\partial z}\\
+\frac{2}{\rho}\frac{\partial}{\partial z}\left(\eta \frac{\partial \varv_z}{\partial z}\right)
+\frac{\epsilon^{2}}{\rho r}\frac{\partial}{\partial r}\left(r \eta \frac{\partial \varv_z}{\partial r}\right)\\
-\frac{2}{3}\frac{\epsilon}{\rho}\frac{\partial}{\partial z}\Bigg[\frac{\eta}{r}\frac{\partial}{\partial r}\Big(r\varv_r\Big) \Bigg]-\frac{2}{3\rho}\frac{\partial}{\partial z}\left(\eta \frac{\partial \varv_z}{\partial z}\right)\\
+\frac{\epsilon}{\rho r} \frac{\partial}{\partial r}\left(\eta r\frac{\partial \varv_r}{\partial z}\right) \ .
\end{aligned}
\end{equation}

\subsubsection*{Order $\epsilon^{0}$:}
\begin{equation}
    0 = -\frac{z}{r^{3}} - n \frac{\partial c_{\rm s{0}}^{2}}{\partial z} - \frac{1}{\gamma 
\tilde{\beta}}\frac{1}{\rho_{0}}\frac{\partial B_{0}^{2}}{\partial z},
\end{equation}
With $ B_0=0$, we have
\begin{equation}
\frac{z}{r^{3}}= - n \frac{\partial c_{\rm s{0}}^{2}}{\partial z},
\label{vertmom0}
\end{equation}
which is the vertical hydrostatic equilibrium equation, with no contribution
from the magnetic pressure.

\subsubsection*{Order $\epsilon^{1}$:}
\begin{equation}
\begin{aligned}
\frac{2}{\gamma\tilde{\beta}}
\left[B_{z0}\frac{\partial B_{z1}}{\partial z}
-\frac{\partial}{\partial z}\left(B_0B_1\right)\right]
-\frac{2}{3r}\frac{\partial}{\partial z}\left[\eta_0
\frac{\partial}{\partial r}\left(r\varv_{r1}\right)\right]\\
+\frac{1}{r}\frac{\partial}{\partial r}
\left(\eta_0r\frac{\partial\varv_{r1}}{\partial z}\right)=0.
\end{aligned}
\label{vertmom1a}
\end{equation}
With $B_{z0}=B_0=0$ we obtain:
\begin{equation}
\begin{aligned}
\frac{\partial}{\partial z}
\left[\eta_0\frac{\partial}{\partial r}\left(r\varv_{r1}\right)\right]
=\frac{3}{2}\frac{\partial}{\partial r}
\left(\eta_0r\frac{\partial\varv_{r1}}{\partial z}\right).
\end{aligned}
\label{vertmom1}
\end{equation}

\section{Numerical setup}
%+++++++++++++++++++++++++++++++++++++++++++++++++++++++++++++++++++
\begin{figure}
\includegraphics[width=\columnwidth]{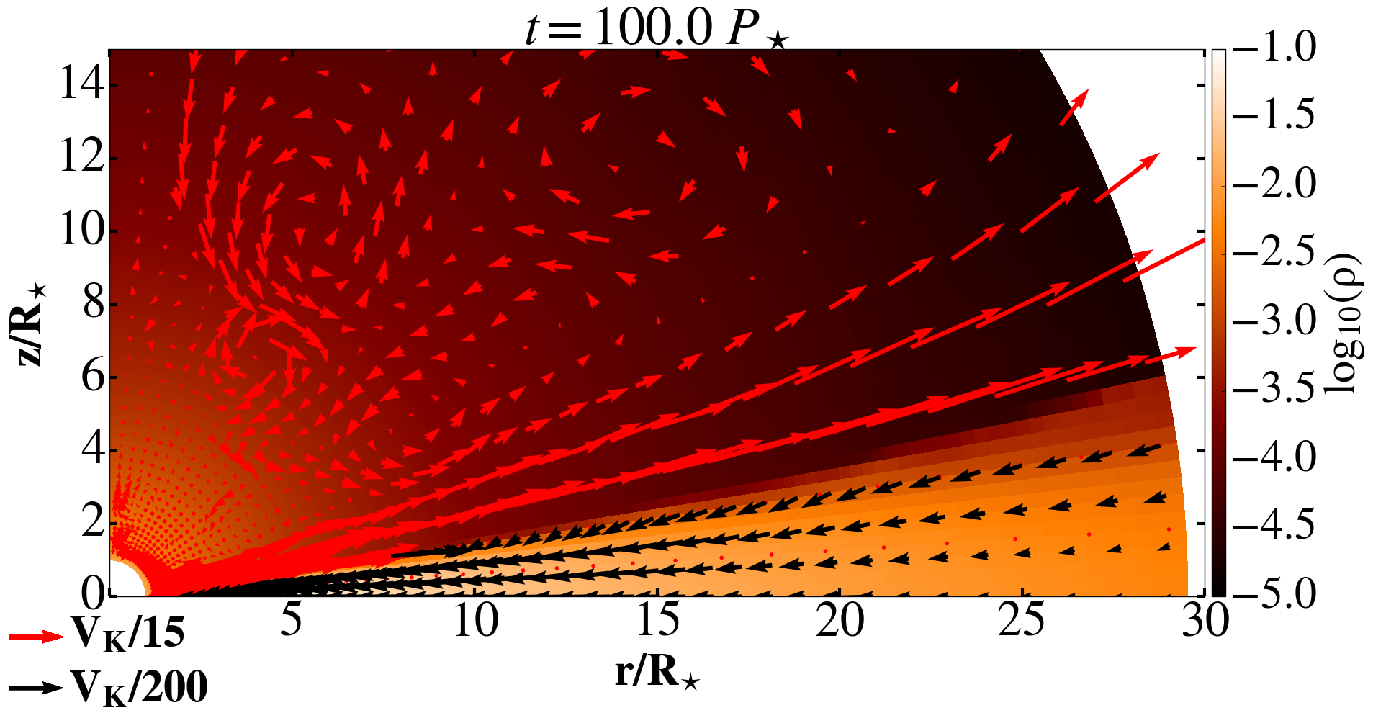}
\caption{Capture into our hydrodynamic simulation after T=100
stellar rotations. The matter density is shown in logarithmic
color grading in code units, with a sample of velocity vectors.
Colors and vectors have the same meaning as in Fig.~\ref{fig:diskreg}.
}
\label{fig:hdsol}
\end{figure}
%+++++++++++++++++++++++++++++++++++++++++++++++++++++++++++++++++++
\begin{figure}
\includegraphics[width=\columnwidth]{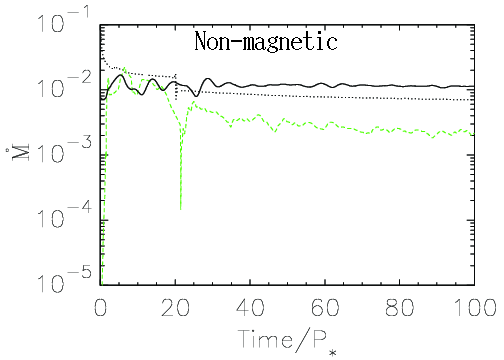}
\caption{Illustration of the quasi-stationarity of our solution in the
HD case, with the mass flux computed in the code units
$\dot{M}_0=\rho_{\mathrm d0} R_\star^2 v_{\mathrm K\star}=\rho_{\mathrm d0} \sqrt{GM_\star R_\star^3}$.
Evolution in time of the mass flux through
the disk, onto the star, and in the polar wind is shown with the solid, dot-dashed, and dashed (green) lines, respectively.
}
\label{fig:maccmominthd}
\end{figure}
%+++++++++++++++++++++++++++++++++++++++++++++++++++++++++++++++++++
%++++++++++++++++++++++++++++++++++++++++++++++++++++++++++++++++++++++
\begin{figure}
\includegraphics[width=\columnwidth]{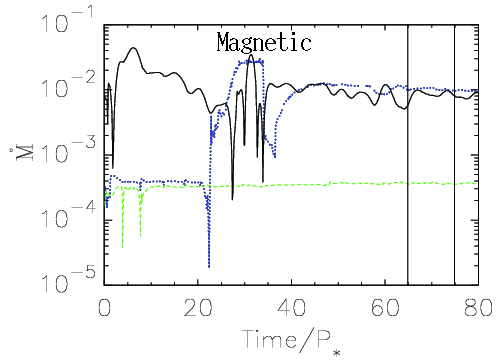}
\caption{Illustration of quasi-stationarity in our magnetic case
solutions, in the same (code) units as in Fig.~\ref{fig:maccmominthd}. 
Shown is the mass flux in the various components of the flow
through the disk at r=15$R_\star$ (solid black line), onto the stellar
surface (dotted blue line), and into the magnetospheric wind (dashed green line).
Vertical thin lines mark the interval in which we take
the average for quasi-stationary solution. 
}
\label{fig:maccmomint1}
\end{figure}
%++++++++++++++++++++++++++++++++++++++++++++++++++++++++++++++++++++++
%+++++++++++++++++++++++++++++++++++++++++++++++++++++++++++++++++++
\begin{figure}
\includegraphics[width=0.9\columnwidth]{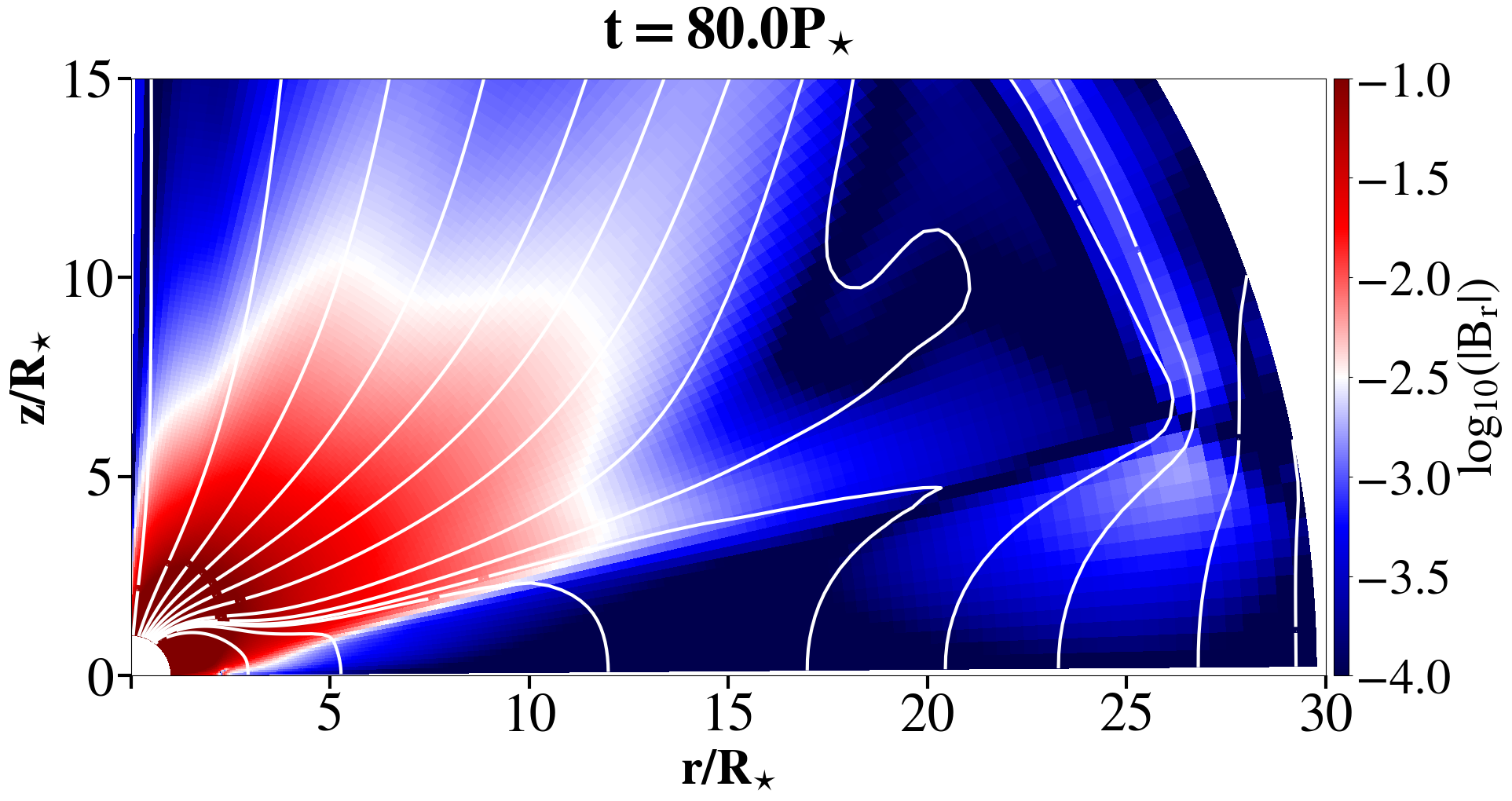}
\includegraphics[width=0.9\columnwidth]{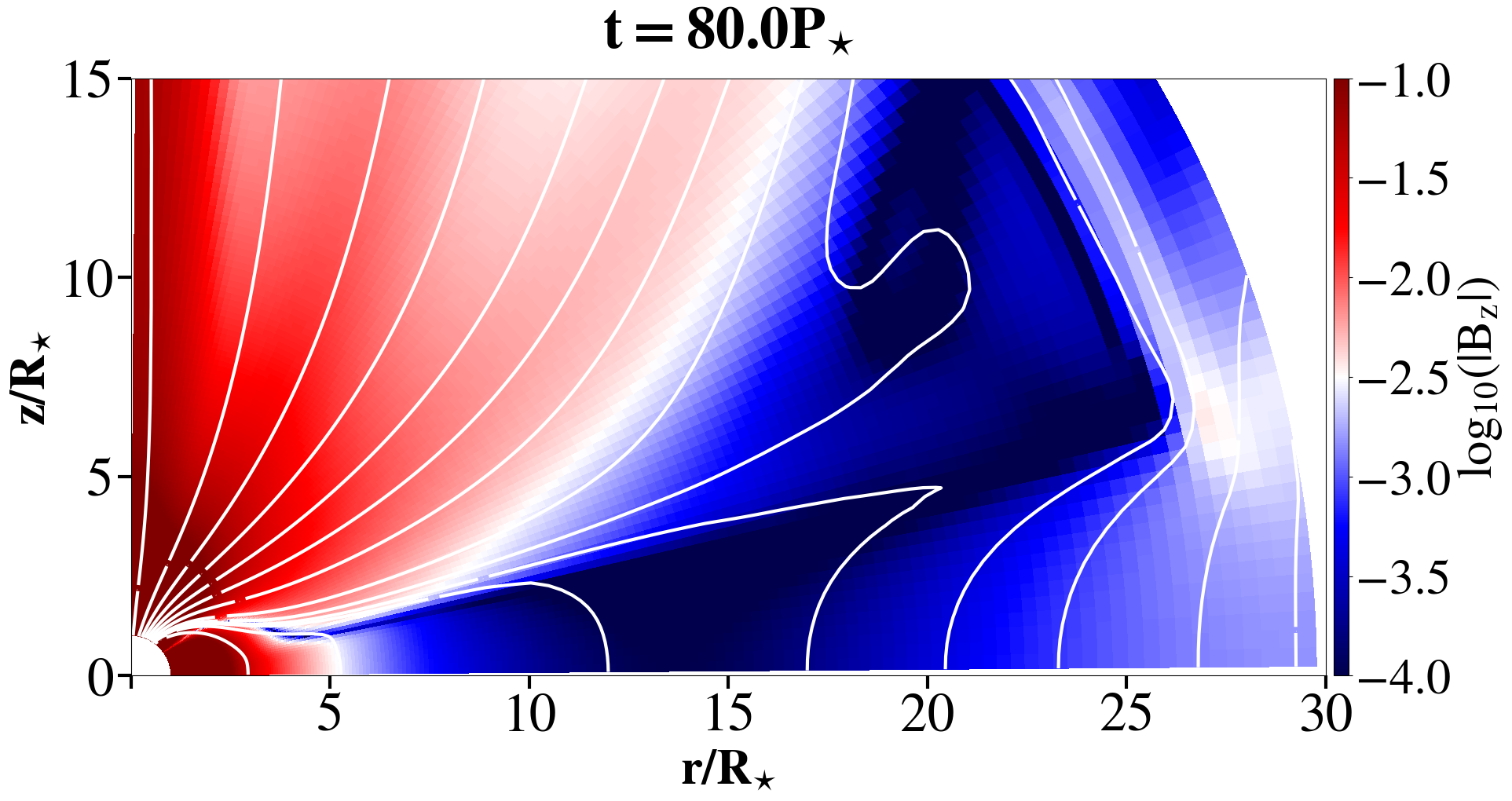}
\includegraphics[width=0.9\columnwidth]{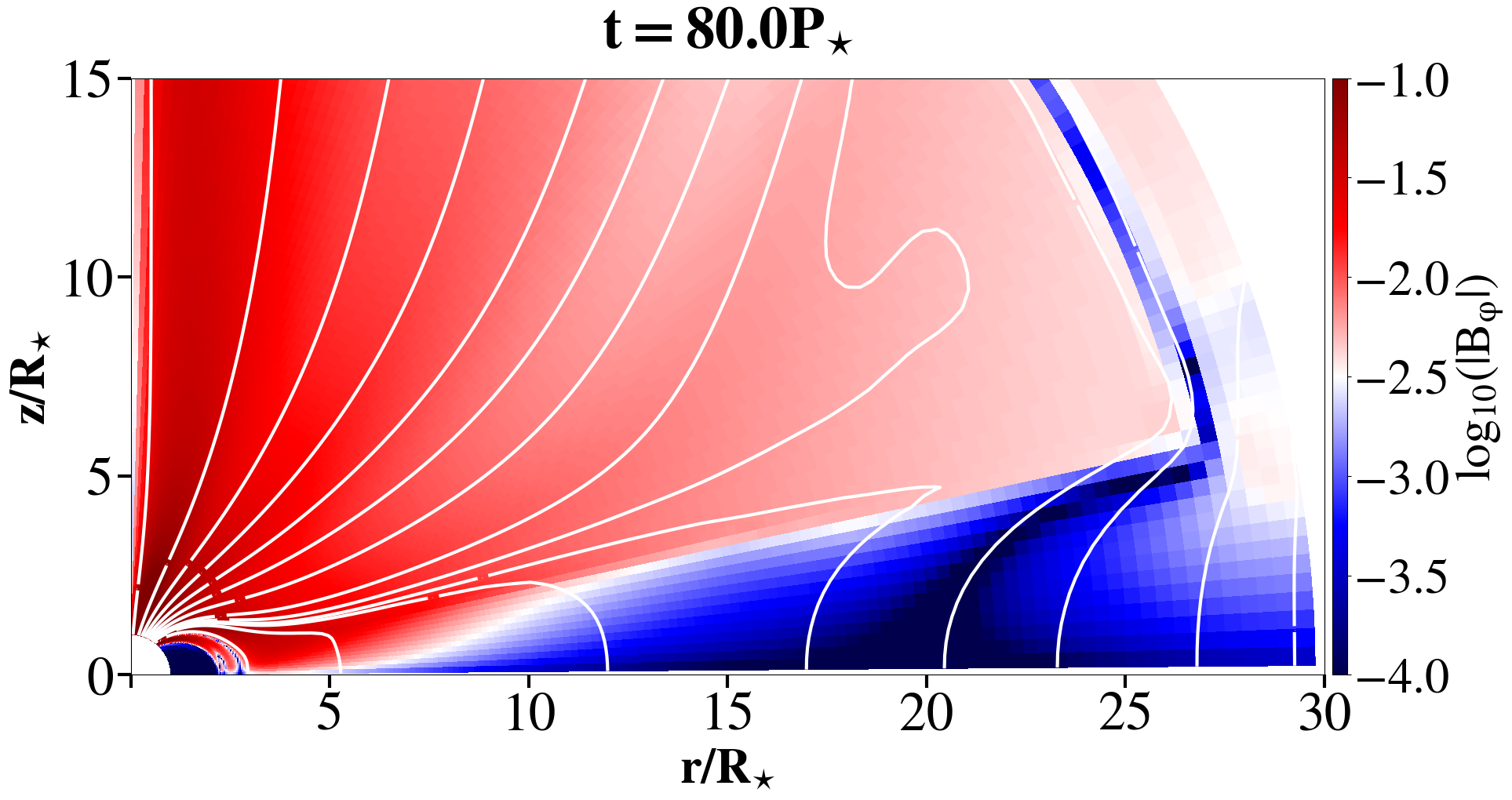}
\caption{Cylindrical components $B_r$, $B_z$, and $B_\varphi$ of the
magnetic field in our simulation from Fig.~\ref{fig:diskreg}, shown
with logarithmic color grading in code units. The total field is
shown in Fig.~\ref{btotcomp}. In the middle part of the disk, the
magnetic field is orders of magnitude smaller than in the
corona, as predicted by our analytical solution in the disk. Magnetic
field is dominant outside of the disk, inside the disk it hardly influences the flow.
}
\label{btotcomp2}
\end{figure}
\begin{figure}
\includegraphics[width=0.9\columnwidth]{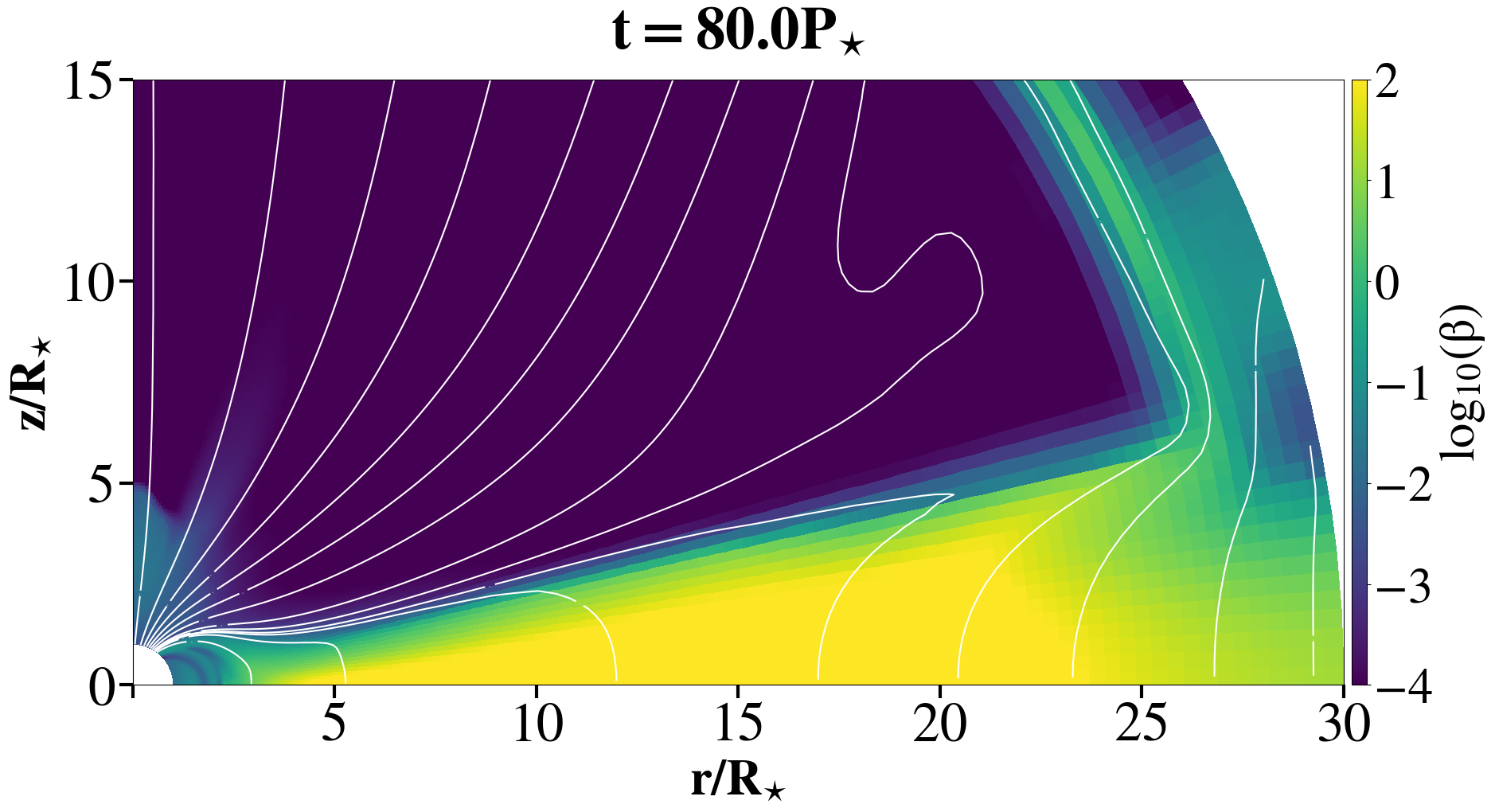}
\caption{Plasma $\beta=8\pi P/B_p^2$ in our simulation from Fig.~\ref{fig:diskreg}, shown
with logarithmic color grading in code units. In the middle part of the disk, where $\beta>>1$, the magnetic field does not control the flow.
}
\label{plasbeta}
\end{figure}
%%+++++++++++++++++++++++++++++++++++++++++++++++++++++++++++++++++++
In \cite{cem19}, following \cite{zf09} we solve the non-ideal MHD
equations using the {\sc pluto} (v.4.1) code \citep{m07} in
the spherical grid. The resolution is $R\times\theta=(217\times 100)$
grid cells, in a logarithmically stretched radial grid and in the first quarter of the meridional plane 
%a half of the meridional half-plane 
in a uniform polar-angle grid, $\theta\in[0,\pi/2]$.
In our numerical upper half-plane simulation we took reflective boundary
conditions at $z=0$, in agreement with reflection symmetry in
the equatorial plane.

The grid resolution is chosen so that the accretion column width is
captured with more than 10 grid cells, and radially logarithmic stretched
grid gives the largest precision where it is needed, close to the disk inner
radius. Simulations with half this resolution in the latitudinal direction
and a corresponding logarithmic spacing, $R\times\theta=(109\times 50)$ (to
maintain the square shape of the computation grid cells) give qualitatively
similar results. In simulations with  radial extension
of the computational box increased 10-fold, up to $R_\mathrm{max}=300~R_\star$, where we
performed long lasting simulations, for up to 5000$P_\star$, the results are
also qualitatively similar to the results obtained in the higher resolution runs reported here.

The viscosity and resistivity are parameterized by the
\citet{ss73} $\alpha$-prescription
as proportional to $c_{\mathrm s}^2/\Omega_{\mathrm K}$. For the
magnetic field, a split-field method is used, so that we evolve in
time only changes from the initial stellar magnetic
field \citep{tan94,pow99}, with the constrained transport
method used to maintain $\nabla\cdot{\mathbf B}=0$. Simulations were
performed using the second-order piecewise linear reconstruction and an
approximate Roe solver. The second-order time-stepping (RK2) was employed.

As an initial condition we use the HD solution of a KK00 disk, adding a
hydrostatic corona and the stellar magnetic dipole field atop a rotating
stellar surface. We derive the custom boundary condition for
$B_\varphi$ from the condition for the stellar surface as a rotating
perfect conductor. During the simulation the corona and the magnetic
field evolve, exerting pressure on the disk surface.

Our computational domain reaches into the middle disk region, 
where the resistivity adds to the viscosity as a dissipation mechanism.
This could make some of the assumptions from the purely HD disk
implausible---with the help of numerical simulations we check whether
or not this is true. A capture in our HD solution after 100 stellar
rotations is shown in Fig.~\ref{fig:hdsol}. In this case, the accretion
onto the star goes through the disk connected to the stellar equator. The
mass flux onto the star and into the stellar wind during the simulation
are shown in Fig.~\ref{fig:maccmominthd}.

Here we present the results in our viscous HD and non-ideal MHD numerical
simulations in the physical domain reaching 30 stellar radii,
$R_\mathrm{max}$=30$R_\star$, with the stellar rotation rate
$\Omega_\star=0.20~\Omega_\mathrm{ms}$, at 20\% of the equatorial
mass-shedding (``break-up'') limit, equal to the Keplerian angular
velocity at the stellar equator
$\Omega_\mathrm{ms}=\sqrt{GM_\star/R_\star^3}$. Thus, the corotation
radius is $R_{\rm{cor}}=(GM_\star/\Omega_\star^2)^{1/3}=(0.20)^{-2/3}
R_\star\approx2.9R_\star$. The anomalous viscosity parameter (much
larger than their microscopic equivalent, assuming that dissipation
is a result of turbulence) is taken to be $\alpha_{\rm v}=1$. It is a
free parameter in the simulation. In the magnetic case we add the stellar
dipole field and the resistivity parameter $\alpha_{\rm m}=1$, so that the
magnetic Prandtl number $P_{\rm m}=2\alpha_{\rm v}/(3\alpha_{\rm m})$=0.67.

In Fig.~\ref{btotcomp2} are shown the components of the magnetic field, to
illustrate the outcome summarized in Fig.~\ref{btotcomp} that magnetic field
does not influence the solution in the disk in the lowest, zeroth order. The large value of plasma-$\beta$ in the disk middle part, as shown in Fig.~\ref{plasbeta}, also supports this conclusion.

These simulations can be applied to any of a number of objects. Here we
are interested in neutron stars in X-ray pulsars. Much higher field values,
as would be needed for the classic X-ray pulsars such as Her X-1, would
require a much shorter time-step and make the simulation prohibitively
expensive. As already remarked, one of the motivations of our analytic
study is to confidently allow our results to be scaled to stronger magnetic
fields.

For neutron stars, with mass of $M_\star$=1.4$M_\sun$, radius
$R_\star$=10~km and with the disk density unit chosen as
$\rho_{\mathrm d0}$=$5\times10^{-6}$g/cm$^3$, from
$\dot{M}_0$=$\rho_{\rm d0}R_\star^2v_{\rm K\star}$ we have
$\dot{M_0}$=5.7$\times10^{-7}~M_\sun$/yr as our unit of mass accretion.
The initial coronal density was defined as a free parameter in the code,
$\rho_{\rm c0}$=0.01$\rho_{\rm d0}$. The unit of the scaled magnetic field is then
$\tilde{B}$=$v_{\rm K\star}\sqrt{\rho_{\rm d0}}$=2.93$\times10^7\,$G. The free
parameter $\mu$=$\sqrt{4\pi}$(0.35, 0.7, 1.05, 1.4) for the magnetic field
strength\footnote{The factor $1/\sqrt{4\pi}$ is absorbed in $B$ in the
PLUTO code.} gives the stellar field
$B_\star$=(0.36, 0.73, 1.09, 1.45)$\times10^8$~G, close to the accepted
value for millisecond accreting pulsars. In the Classical T-Tauri star
cases, the stellar mass is $M_\star$=0.5$M_\sun$, radius $R_\star$=2$R_\sun$,
and we choose the stellar angular velocity of $0.2\,\Omega_{\mathrm ms}$,
which gives the stellar rotation period of
$P_\star$=2$\pi/\Omega_\star$=2.32 days. If we choose the same mass accretion
unit as above, we have $\rho_{\mathrm d0}\sim8.5\times10^{-11}$g/cm$^3$, and 
$\tilde{B}$=$v_{\rm K\star}\sqrt{\rho_{\rm d0}}\simeq$0.2~kG, and the stellar fields
with the same free parameters $\mu$ as above would produce
$B_\star\simeq$(0.25, 0.5, 0.75, 1)~kG, realistic for the YSO cases. A table
for rescaling to different types of objects is given in \cite{cem19}, where
a parameter study was performed with the same set-up.
\end{appendix}%closing of Appendix
\end{document}